\documentclass[12pt]{article} \input epsf.tex
\usepackage{graphicx}
\usepackage[tbtags,fleqn]{amsmath}

\newcommand{\be}{\begin{equation}}
\newcommand{\ee}{\end{equation}}
\newcommand{\bear}{\begin{eqnarray}}
\newcommand{\eear}{\end{eqnarray}}
\newcommand{\ba}{\begin{array}}
\newcommand{\ea}{\end{array}}
\newcommand{\lae}{\begin{array}{c}\,\sim\vspace{-21pt}\\<
\end{array}}
\newcommand{\gae}{\begin{array}{c}\,\sim\vspace{-21pt}\\>
\end{array}}
\begin{document}

\pagestyle{empty} \begin{titlepage}
\def\thepage {} 

\title{\Large \bf
$Z^\prime$ Gauge Bosons at the Tevatron
}

\author{\normalsize
\bf \hspace*{-.3cm} Marcela Carena$^1$, Alejandro Daleo$^{1,2}$
Bogdan A.~Dobrescu$^1$, Tim~M.P.~Tait$^1$
\\ \\ 
{\small {\it
$^1$ Theoretical Physics Department, Fermilab, Batavia, IL 60510, USA }}\\
{\small {\it 
$^2$ Departamento de F\'{\i}sica, Universidad Nacional de La Plata, }}\\
{\small {\it C.C. 67-1900 La Plata, Argentina.}}\\
}

\date{\small August 5, 2004 } \maketitle

\vspace*{-8.4cm}
\noindent \makebox[11cm][l]{\small \hspace*{-.2cm} } {\small  FERMILAB-Pub-04/129-T} \\
\makebox[11cm][l]{\small \hspace*{-.2cm} } {\small hep-ph/0408098 } {} \\
{\small } \\

\vspace*{7.9cm}

\begin{abstract}
We study the discovery potential of the Tevatron for a $Z^\prime$ gauge boson.
We introduce a parametrization of the $Z^\prime$ signal 
which provides a convenient bridge between collider searches and
specific $Z^\prime$ models. 
The cross section for 
$p \overline{p} \rightarrow Z^\prime X \rightarrow \ell^+ \ell^- X$ 
depends primarily on the $Z^\prime$ mass and
the $Z^\prime$ decay branching fraction into leptons times the 
average square coupling to up and down quarks.
If the quark and lepton masses are generated as in the standard model, 
then the $Z^\prime$ bosons accessible at the Tevatron 
must couple to fermions proportionally to 
a linear combination of baryon and lepton numbers in order to
avoid the limits on $Z-Z^\prime$ mixing.
More generally, we present several families of $U(1)$ 
extensions of the standard
model that include as special cases many of the 
$Z^\prime$ models discussed in the literature. 
Typically, the CDF and D0 experiments are expected to probe $Z^\prime$-fermion 
couplings down to 0.1 for $Z^\prime$ masses in the 500--800 GeV range, 
which in various 
models would substantially improve the limits set by the LEP experiments.

\end{abstract}

\vfill \end{titlepage}

\baselineskip=18pt \pagestyle{plain} \setcounter{page}{1}
\section{Introduction} \setcounter{equation}{0}

An important question in particle physics today is whether there are any new
gauge bosons beyond the ones associated with the $SU(3)_C \times SU(2)_W \times U(1)_Y$ 
gauge group. This question is interesting by itself, given that the 
selection of the gauge bosons observed so far remains mysterious. Furthermore,
new gauge bosons are predicted within many theories beyond the 
Standard Model (SM) which have been developed to provide answers to its many
open questions.

The simplest way of extending the SM gauge structure is to include a 
second $U(1)$ group. The associated gauge boson, usually labeled $Z^\prime$,
is an electrically-neutral spin-1 particle.
If the new gauge coupling is not much smaller than unity, then the 
$U(1)$ group must be spontaneously broken at a scale larger than 
the electroweak scale in order to account for the nonobservation of the 
$Z^\prime$ boson at LEP and run I of the Tevatron. 
In this article, we study the $Z^\prime$ discovery potential of the run II
of the Tevatron, the highest energy hadron machine operating for the
next few years. 

The theoretical framework for studying $Z^\prime$ production at hadron colliders 
has been developed more than two decades ago \cite{Eichten:1984eu}. Nevertheless, 
various pieces of information collected recently have an impact on our 
attempt of addressing a number of specific questions:
What $Z^\prime$ parameters are relevant for Tevatron searches? What 
regions of the parameter space are not ruled out by the LEP experiments, and 
would allow a $Z^\prime$ discovery at the Tevatron? In case of a discovery, how 
can one differentiate between the models that may accommodate a $Z^\prime$
boson? 

It is often assumed that the $Z^\prime$ couplings have certain values motivated 
by some narrow theoretical assumptions, allowing for the derivation of a $Z^\prime$
mass bound  \cite{Hewett:1988xc, Cvetic:1995rj}. 
The opposite approach of leaving the couplings arbitrary \cite{Leike:1998wr}
suffers from the existence of too many free parameters. However, a few
theoretical constraints are sufficiently generic so that it is reasonable to focus
on the region of the parameter space that satisfies them.
This observation, used to define the 
so-called nonexotic $Z^\prime$ bosons \cite{Appelquist:2002mw}, 
underscores the importance of the 
$Z^\prime$ couplings to the SM fermions for collider phenomenology 
\cite{Freitas:2004hq}, while reducing the set of $Z^\prime$ parameters.

In this article, we address $Z^\prime$ models both from a theoretical
perspective and with respect to their potential 
observation at hadron colliders.
In Section~\ref{sec:params} we  present the theoretical framework needed to
describe a new neutral gauge boson.
We analyze the constraints due to gauge anomaly cancellation
and the gauge invariance of the quark and lepton Yukawa couplings, 
and discuss what new physics would soften these constraints.
We identify several interesting families of $Z^\prime$ models, and then 
derive the LEP limits.
Section~\ref{sec:tevatron} 
is concerned with $Z^\prime$ production at hadron colliders,
including a survey of theoretical tools to describe $Z^\prime$ events, and
a convenient parameterization of limits from searches that simplifies
comparison of experimental results with theoretical models.
Section 4 summarizes our conclusions.

\section{Parameters Describing New Neutral Gauge Bosons}
\label{sec:params}

Any new gauge boson is characterized by a mass and a number of 
coupling constants. All these parameters appear in the Lagrangian, 
which is constrained by gauge and Lorentz invariance.
In this section we present a theoretical framework that is sufficiently  
general to account for the parameters that are relevant for 
$Z^\prime$ searches at the Tevatron.
We discuss the theoretical constraints within realistic extensions of 
the SM, and then discuss the LEP limits.

\subsection{$Z^\prime$ mass and $Z-Z^\prime$ mixing}
\label{subsec:zpmass}

Consider the SM gauge symmetry extended by 
one Abelian gauge group, $SU(3)_C\times SU(2)_W \times U(1)_Y 
\times U(1)_z$. 
The scalar sector responsible for the spontaneous breaking
of the gauge symmetry down to $SU(3)_C\times U(1)_{\it em}$
includes at least one Higgs doublet and an $SU(2)_W$ singlet, $\phi$,
with VEV $v_\phi$.
As we will explicitly show below, the constraints on the interactions of the 
$U(1)_z$ gauge boson with quarks and leptons are relaxed in the presence
of two Higgs doublets, $H_1$ and $H_2$, with aligned VEVs $v_{H_1}$ 
and $v_{H_2}$. To be general, we will concentrate on this case 
in what follows.
The hypercharges of $H_1$, $H_2$, and $\phi$ are given by 
$+1$, $+1$ and $0$, respectively, so that electric  charge
is conserved.

In a basis where the three electrically neutral gauge bosons, 
$W^{3 \mu}$, $B_Y^\mu$ and $B_Z^\mu$, have diagonal kinetic terms,
their mass terms are given by:
\be
\frac{v_{H_1}^2}{8} \left(g W^{3 \mu} - g_Y B_Y^\mu - z_{H_1} 
g_z B_z^\mu \right)^2
+\frac{v_{H_2}^2}{8} \left(g W^{3 \mu} - g_Y B_Y^\mu - z_{H_2} 
g_z B_z^\mu \right)^2
+\frac{v_\phi^2}{8} \left( z_\phi g_z B_z^\mu  \right)^2 ~,
\ee
where $g, g_Y, g_z$ are the $SU(2)_W \times U(1)_Y \times U(1)_z$
gauge couplings, and the weak mixing angle is 
given as usual by $\tan\theta_w = g_Y/g$.
The diagonalization of these mass terms yields the three physical states,
the photon (labeled by $A^\mu$), the observed $Z$ boson, and 
the hypothetical $Z^\prime$ boson:
\bear
\label{eq:states}
A_\mu & = & W^{3 \mu} \sin\theta_w + B_Y^\mu \cos\theta_w ~ , 
\nonumber \\ [2mm]
Z_\mu & =  & W^{3 \mu} \cos\theta_w -  B_Y^\mu \sin\theta_w + 
\epsilon B_z^\mu ~ , 
\nonumber \\ [2mm]
Z^\prime_\mu & = &   B_z^\mu - \epsilon 
\left( W^{3 \mu} \cos\theta_w -  B_Y^\mu \sin\theta_w \right) ~ .
\eear
To obtain this result we have ignored terms of order $\epsilon^2$, where 
$\epsilon$ is the mixing angle between the SM 
$Z$ boson and $B_z^\mu$,
\bear\label{eq:angle}
\epsilon = \frac{\delta M^2_{Z Z^\prime} }{M^2_{Z^\prime} - M^2_Z} ~.
\eear
The mass-squared parameters introduced here are related to the VEVs by
\bear
\label{eq:bosonmasses}
M^2_{Z} & =  & \frac{ g^2 }{4 \cos^2\theta_w} 
\left( v_{H_1}^2 + v_{H_2}^2  \right)
\left[ 1 + O\left( \epsilon^2 \right) \right]~,
\nonumber \\ [2mm]
M^2_{Z^\prime} & =  & \frac{ g_z^2 }{4} 
\left(z_{H_1}^2  v_{H_1}^2 +  z_{H_2}^2  v_{H_2}^2 + z_\phi^2 v_\phi^2 \right) 
\left[ 1 + O\left( \epsilon^2 \right) \right]~,
\nonumber \\ [2mm]
\delta M^2_{Z Z^\prime} & =  & - \frac{ g g_z}{4 \cos\theta_w} 
\left( z_{H_1} v_{H_1}^2 +  z_{H_2} v_{H_2}^2 \right) ~.
\eear
$M_{Z}$ and $M_{Z^\prime}$  are the physical masses of the neutral 
gauge bosons up to corrections of order $\epsilon^2$.

A $Z^\prime$ boson which mixes with the SM $Z$
distorts its properties, such as couplings to fermions and mass relative to
electroweak inputs. 
Precision measurements of observables,
mostly on the $Z$ pole at LEP I and SLC, 
have verified the SM $Z$ properties at or below the per mil
level\footnote{The notable exception is $\sin^2 \theta_W$ from hadronic $Z$ 
decays, which deviates at the few $\sigma$ level \cite{Altarelli:2004fq}, 
and plays an interesting role in the fit to the SM Higgs mass 
\cite{Chanowitz:2001bv}.  These deviations
have been argued as evidence for the existence of a $Z^\prime$
with non-universal interactions \cite{Erler:1999nx}; we
shall not pursue this line of reasoning here.} \cite{Altarelli:2004fq},
imposing a severe upper bound \cite{Abreu:1994ri}
on the mixing angle between the 
$Z$ and $Z^\prime$: $|\epsilon| \lae 10^{-3}$.
Therefore, it is justified to treat the mixing as a perturbation as was
done above.


From Eqs.~(\ref{eq:angle}) and (\ref{eq:bosonmasses}) it
follows that the mixing angle is given by 
\bear
\label{eq:mix}
|\epsilon | \approx \frac{ g_z}{ g} 
\left(\frac{\cos\theta_w}{M^2_{Z^\prime}/M^2_Z -1} \right) 
\frac{ \left| z_{H_1} + z_{H_2} \tan^2\!\beta \right| }{ 1 + \tan^2\!\beta }
\eear
where $\tan\beta = v_{H_2}/ v_{H_1}$. 
At least one of $v_{H_1}$ or $ v_{H_2}$ has to be of the order of the 
electroweak scale (to generate $M_W$, $M_Z$ and $m_t$ appropriately), 
so without loss of generality we can set 
$v_{H_2} \sim O(246)$ GeV.
Therefore, $\tan\beta \gae O(1)$.
Normalizing the largest quark $U(1)_z$ charges to be of order unity,
the $Z^\prime$ production at the Tevatron is sizable only if 
the gauge coupling $g_z$ is not much smaller than unity.
The mass range typically interesting at the Tevatron
is roughly 0.2 TeV $ < M_{Z^\prime} < $ 0.7 TeV.
Based on these considerations we find that the order of magnitude of 
the mixing angle is given by
$\epsilon \sim (z_{H_1}\cot^2\!\beta + z_{H_2})  M^2_Z/ M^2_{Z^\prime}$.
The constraint $|\epsilon| \lae 10^{-3}$ implies 
$z_{H_1}\cot^2\!\beta + z_{H_2} \ll 1$. 
Although $\tan^2\beta$ could be close to $-z_{H_1}/z_{H_2}$,
this would be a fine tuning, because  the value of $\tan\beta$ is set 
by the Higgs masses and self-interactions,
and has no reason to be related to the ratio of Higgs charges.
Therefore, in the absence of fine-tuning, 
a $Z^\prime$ accessible at the Tevatron requires  
$|z_{H_2}| \ll 1$ and either $|z_{H_1}|\ll 1$
or $\tan\beta \gg 1$.
It is usually expected that the charges of various fields are either 
all of the same order or vanish.  Although exceptions exist, such as 
extra dimensional models with brane kinetic terms \cite{Carena:2002me}
on the Higgs brane, which motivate much a smaller effective charge for
the Higgs than for the (bulk) fermions, 
we restrict attention here to the following two cases:
\bear 
z_{H_2} = 0  \;\; {\rm and} \; \; z_{H_1} = 0
\eear
or 
\bear 
z_{H_2} = 0 \;\; {\rm and} \; \;\tan\beta \gae 10 ~.
\eear

\subsection{Couplings to fermions}
\label{subsec:zpcoup}

The renormalizable
interactions of the $Z^\prime$ boson with the SM fermions 
are described by the following terms in the Lagrangian density:
\bear
\sum_{f} z_f g_z Z^\prime_\mu \overline{f} \gamma^\mu f ~,
\eear
where $f = e_R^j, l_L^j, u_R^j, d_R^j, q_L^j$ are the 
usual lepton and quark fields in the weak eigenstate basis; 
$l_L^j=(\nu_L^j, e_L^j)$ and $q_L^j=(u_L^j, d_L^j)$ are the $SU(2)_W$ doublet 
fermions.
The index $j$ labels the three fermion generations. 
Altogether there are 15 fermion charges, $z_f$.

The observed quark and lepton masses and mixings restrict these 
fermion charges, so that certain gauge and Lorentz invariant terms 
can appear in the Lagrangian.
In the SM, the terms responsible for the charged 
fermion masses are 
\bear \label{eq:masses}
\lambda_{jk}^d \, \overline{q}_L^j d_R^k H +
\lambda_{jk}^u \, \overline{q}_L^j u_R^k i\sigma_2 H^\dagger + 
\lambda_{jk}^e \, \overline{l}_L^j e_R^k H  + {\rm h.c.} ~,
\eear 
where $j,k = 1,2,3$ label the fermion generations, and 
$\lambda_{jk}^d$, $\lambda_{jk}^u$, $\lambda_{jk}^e$ are Yukawa couplings.
In the two Higgs-doublet model described above, the Higgs doublet $H$ is 
replaced by linear combinations of $H_1$ and $H_2$. 

As discussed in section 2.1, the $Z^\prime$ bosons relevant for 
Tevatron searches
have small mixing with the $Z$ boson, which effectively implies that any 
Higgs doublet with a VEV of order the electroweak scale is neutral under 
the $U(1)_z$ symmetry.
In particular, if only one Higgs doublet is present, then 
its $U(1)_z$ charge would have to vanish. 
Given that the total charge of the quark mass terms shown in 
Eq.~(\ref{eq:masses}) has to be zero, the quark masses and CKM elements 
may then be accommodated only if the quarks have generation independent 
$U(1)_z$ charges, and $z_u = z_d = z_q$,
where  $z_u$ and $z_d$ are the right-handed up- and down-type quark charges, 
and $z_q$ is the left-handed quark doublet charge.
One may relax this condition in the two Higgs-doublet model if, for example, 
$H_2$ couples to the up-type quarks,  while $H_1$ couples 
to the down-type quarks, has nonzero charge, and $\tan\beta$ is large.
In that case $z_d$ may be different from $z_u$ and $z_q$, 
but one still needs to impose
\bear \label{eq:charge}
z_u = z_q ~,
\eear
so that the large top-quark mass may be generated.

We emphasize that we have derived this strong conclusion based on 
reasonable but not infallible arguments. One loophole is that 
some of the terms in Eq.~(\ref{eq:masses}) may be replaced by 
higher-dimensional operators such as 
$\overline{q}_L^j d_R^k H (\phi /M_{\rm heavy})^p$, where $p$ is an integer
and $M_{\rm heavy}$ is the mass scale where this dimension-$(4+p)$
operator is generated. Since the weak-singlet scalar $\phi$ 
has a nonzero charge under $U(1)_z$,
the relations between the various quark charges may be changed.
The higher-dimensional operators may be induced in a renormalizable
quantum field theory by the exchange of heavy fermions
that have Yukawa couplings to both the Higgs doublets and $\phi$.
Another loophole is that both Higgs doublets may be charged 
under  $U(1)_z$ if there is a fine-tuning as discussed in section 2.1,
so that the restrictions on quark charges would be again modified.
Based on these considerations, we will study in some detail the 
implications of Eq.~(\ref{eq:charge}), but we will also consider 
departures from it. 

We point out that generation dependent 
quark charges lead to flavor-changing couplings of the $Z^\prime$
in the mass eigenstate basis, where the fermion mass matrices are diagonal.
Various experimental constraints from 
flavor-changing neutral current processes impose severe constraints
on such  flavor-changing $Z^\prime$ couplings, 
unless $M_{Z^\prime}$ is so large that 
the effects of such a $Z^\prime$ would be beyond the reach of 
even the LHC.  
To avoid these complications we will avoid 
generation dependent quark charges in this paper. In practice this 
does not restrict significantly the generality of our results
because  the Tevatron is typically not very sensitive to $Z^\prime$ 
decaying into quarks, and the production cross section depends
only on an average quark charge.
Therefore, altogether there are three quark charges relevant in what follows:
$z_u, z_d, z_q$.

The masses of the electrically-charged leptons can be induced by the last 
term shown in Eq.~(\ref{eq:masses}) even if the lepton $z$-charges 
are generation dependent. Moreover, no flavor-changing neutral currents are 
induced in the lepton sector by $Z^\prime$ exchange.
The lepton mass terms impose, though, a relation between the left- and right-handed 
lepton $z$-charges: 
$z_{l_j} = z_{e_j}$, $j = 1,2,3$, 
or $z_{l_j} - z_{e_j} = z_{H_1}$ in the two-Higgs doublet model with large 
$\tan\beta$. 
As in the case of the quarks, we allow for 
deviations from these equalities motivated by lepton mass generation via 
higher-dimensional operators. Thus, all six lepton charges, $z_{l_j},z_{e_j}$, 
$j = 1,2,3$ could be relevant for Tevatron studies. 

Additional constraints arise due to the requirement of 
generating neutrino masses and mixings.
The terms in the Lagrangian responsible for these are given by 
\bear\label{eq:neutrino}
\frac{c_{jk}}{M_\nu}\, \overline{l^c}_L^j l_L^k \, H^\top i\sigma_2 H 
+ \lambda_{jk^\prime}^\nu \, 
\overline{l}_L^j i\sigma_2 \nu_R^{k^\prime} H^\dagger 
+ m_{j^\prime k^\prime}^\nu \overline{\nu^c}_R^{j^\prime} \nu_R^{k^\prime} 
+ {\rm h.c.} ~,
\eear 
where we have included right-handed neutrinos, $\nu_R^{j^\prime}$, 
which are singlets under the SM gauge group. If these 
are not present, 
then the last two terms in the above equation vanish.
If there are $n$ right-handed neutrino flavors, 
then $j^\prime, k^\prime = 1, ..., n$.
For $n \ge 2$ all dimensionless coefficients $c_{jk}$ of the
above lepton-number-violating terms may vanish. 
The other parameters appearing in Eq.~(\ref{eq:neutrino}) are as follows:
$M_\nu$ is the mass scale where the 
lepton-number-violating terms are generated,
$m_{j^\prime k^\prime}^\nu$ are right-handed neutrino Majorana masses,
$\lambda_{jk^\prime}^\nu$ are some Yukawa couplings. 

The requirement that the three active neutrinos mix, so that the 
observed neutrino oscillations can be accommodated, implies that 
the lepton charges are generation independent.
However, as in the case of quarks, the terms in Eq.~(\ref{eq:neutrino})
may be replaced by higher-dimensional operators involving powers 
of $\phi /M_{\rm heavy}$. Furthermore, the tiny neutrino masses 
make the existence of such higher-dimensional operators an attractive
possibility \cite{Appelquist:2002mw}.
If several $\phi$ scalars carry different $U(1)_z$ charges,
one could avoid almost entirely the constraints from  neutrino mixing
on lepton charges.

The six lepton charges
determine the leading decay width of the $Z^{\prime}$ into the 
corresponding leptons: 
\bear \label{eq:widths}
\Gamma(Z^{\prime} \rightarrow e^{+}_j e^{-}_j ) & \approx & 
 (z_{l_j}^2 + z_{e_j}^2) \frac{g_z^2}{24 \pi}  M_{Z^\prime} ~,
\nonumber \\ [2mm]
 \Gamma(Z^{\prime} \rightarrow \nu_L \bar{\nu}_L )
& \approx & \left(z_{l_1}^2 + z_{l_2}^2 + z_{l_3}^2\right) 
\frac{g_z^2}{24\pi}  M_{Z^\prime} ~,
\eear
where $e_j = \{ e, \mu, \tau \}$ for $j=1,2,3$.
Similarly, the quarks charges determine the following decay widths 
of the $Z^{\prime}$:
\bear \label{eq:widths-quarks}
 \Gamma(Z^{\prime} \rightarrow {\rm jets})
& \approx & (2z_q^2 + z_u^2 + z_d^2) \frac{g_z^2}{4\pi}  M_{Z^\prime}
\left(1 + \frac{\alpha_s}{\pi} \right) ~,
\nonumber \\ [2mm]
 \Gamma(Z^{\prime} \rightarrow b \bar{b})
& \approx & (z_q^2 + z_d^2) \frac{g_z^2}{8\pi}  M_{Z^\prime} 
\left(1 + \frac{\alpha_s}{\pi} \right) ~,
\nonumber \\ [2mm]
 \Gamma(Z^{\prime} \rightarrow t \bar{t})
& \approx & (z_q^2 + z_u^2 ) \frac{g_z^2}{8\pi}  M_{Z^\prime} 
\left(1 - \frac{m_t^2}{M_{Z^\prime}^2} \right)
\left( 1 - \frac{4m_t^2}{M_{Z^\prime}^2} \right)^{\! 1/2}\nonumber \\ [2mm]
&&  \times
\left[1 + \frac{\alpha_s}{\pi} + 
O\left( \alpha_s m_t^2/M_{Z^\prime}^2 \right) \right] 
\theta \left(M_{Z^\prime} - 2m_t \right)~.
\eear
where ``jets'' refers to hadrons not containing bottom or top quarks and
we have included the leading QCD corrections, but we have 
ignored electroweak corrections and all fermion masses with the 
exception of the top-quark mass, $m_t$. 
Additional decay modes, into pairs of Higgs bosons 
(if $z_{H_1} \neq 0$ or $z_{H_2}\neq 0$),
CP-even components of the $\phi$ scalar, right-handed neutrinos, 
or other new particles 
might be kinematically accessible and large. Therefore, the total decay width,
$\Gamma_{ Z^{\prime}}$, is larger than or equal to the sum of the 
seven decay widths 
shown in Eqs.~(\ref{eq:widths}) and (\ref{eq:widths-quarks}).

Assuming that the decays into particles other than the SM 
fermions are 
either invisible or have negligible branching ratios, 
the $Z^{\prime}$ properties depend primarily on eleven parameters: 
mass ($M_{Z^\prime}$), total width ($\Gamma_{Z^\prime}$), 
and nine fermion couplings 
($z_{e_j}, z_{l_j}, z_q, z_u, z_d$)$\times g_z$.

\subsection{Realistic models}
\label{subsec:anomalies}



So far we have imposed  $SU(3)_C\times SU(2)_W \times U(1)_Y 
\times U(1)_z$ gauge invariance on the Lagrangian. 
Additional restrictions need to be imposed in order to preserve 
gauge invariance in the full quantum field theory: the fermion content of 
the theory has to be such that all gauge anomalies cancel.
In our case, we need to make sure that there are no 
gauge anomalies due to triangle diagrams with gauge bosons as external lines. 


Triangle diagrams involving two gluons or two $SU(2)_W$ gauge bosons,
and one  $U(1)_z$ gauge bosons give rise to the 
$[SU(3)_C]^2U(1)_z$ and $[SU(2)_W]^2U(1)_z$ anomalies:
\bear\label{eq:a33z}
A_{33z} & = &  3 \left( 2z_q - z_u - z_d \right) ~,
\nonumber \\ [2mm]
A_{22z} & = & 9 z_q + \sum_{j=1}^3 z_{l_j}  ~.
\eear
Triangle diagrams involving $U(1)_Y \times U(1)_z$ gauge bosons 
give rise to the $[U(1)_Y]^2 U(1)_z$, $U(1)_Y [U(1)_z]^2$ and 
$[U(1)_z]^3$ anomalies:
\bear
A_{11z} & = & 2z_q - 16 z_u - 4 z_d + 2 \sum_{j=1}^3 
\left( z_{l_j} - 2 z_{e_j} \right) ~,
\nonumber \\ [2mm]
A_{1zz} & = & 6 \left( z_q^2 - 2 z_u^2 + z_d^2 \right) 
- 2 \sum_{j=1}^3 \left( z_{l_j}^2 - z_{e_j}^2 \right) ~,
\nonumber \\ [2mm]
A_{zzz} & = & 9 \left( 2 z_q^3 - z_u^3 - z_d^3\right) + 
\sum_{j=1}^3 \left( 2z_{l_j}^3 - z_{e_j}^3 \right)
- \sum_{i=1}^n z_{\nu_i}^3 ~,
\eear
where we have included $n$ right-handed neutrinos of charges 
$z_{\nu_i}$ under $U(1)_z$. 
Finally, triangle diagrams involving two gravitons and one
$U(1)_z$ gauge boson contribute to the mixed gravitational-$U(1)_z$
anomaly, which makes general coordinate invariance incompatible 
with $U(1)_z$ gauge invariance:
\bear\label{eq:aggz}
A_{GGz} & = & 9 \left( 2z_q - z_u - z_d \right) 
+ \sum_{j=1}^3 \left( 2z_{l_j} - z_{e_j} \right) - \sum_{i=1}^n z_{\nu_i}
\eear

Gauge invariance at quantum level requires that all the anomalies 
listed in Eqs.~(\ref{eq:a33z})-(\ref{eq:aggz}) vanish, or are exactly canceled
by anomalies associated with some new fermions charged under both 
the SM
gauge group and $U(1)_z$. 
The impact of the new fermions on the $Z^\prime$ properties described here 
can be ignored if they are heavier than the $Z^\prime$.
Altogether, there are six equations that restrict the nine $z$-charges of the 
SM fermions.
Finding solutions to this set of equations is a nontrivial task,
especially if one imposes that the charges are rational numbers,
as suggested by grand unified theories. 

The case where the top-quark mass is generated by a Yukawa coupling 
to a Higgs doublet, as in the SM or the 
Minimal Supersymmetric Standard Model (MSSM),
leads to Eq.~(\ref{eq:charge}), in which case the 
$A_{33z}, A_{22z}$ and $ A_{11z}$ anomalies vanish only if  
\bear
&& z_q = z_u = z_d = -\frac{1}{9}\sum_{j=1}^3 z_{l_j}  ~,
\\ [2mm]
&& \sum_{j=1}^3 \left( z_{l_j} - z_{e_j} \right) = 0 ~.
\eear
The remaining anomaly cancellation conditions, in the absence of 
exotic fermions,
take the following form:
\bear 
&& \sum_{j=1}^3 \left( z_{l_j}^2 - z_{e_j}^2 \right) = 0 ~,
\nonumber\\ [2mm]
&& \sum_{j=1}^3 \left( 2z_{l_j}^3 - z_{e_j}^3 \right) = 
\sum_{i=1}^n z_{\nu_i}^3 
\nonumber\\ [2mm]
&& \sum_{i=1}^n z_{\nu_i} = - 9 z_q
\eear
It is hard to find general solutions to this set of equations. A well known
nontrivial solution is $n = 3$ 
and $z_{l_j}= z_{e_j} = z_{\nu_j} = z_l$, $j = 1,2,3$,
which corresponds to the $U(1)_{B-L}$ gauge group. The associated 
gauge boson, $Z_{B-L}$, is an interesting case of a ``non-exotic'' 
$Z^\prime$ \cite{Appelquist:2002mw} relevant
for Tevatron searches. We have found a generalization of this solution
which preserves the $z_{l_j}= z_{e_j} = z_{\nu_j}$ equalities 
within each generation, but have different lepton charges for different 
generations. In particular, the case of $z_{l_1}= 0$ 
is worth special attention because the $Z^\prime$ does not couple to electrons,
so there are no tight limits from LEP. 
In this case there are only two independent
charges: $z_q$ which sets all quark charges, and $z_{l_2}$ which sets the charges of 
the muon and second-generation neutrinos. Normalizing the gauge coupling such that
$z_q = 1/3$, the $\tau$ and third-generation neutrinos
have charge 
\bear 
z_{l_3} = - 3 - z_{l_2}~.
\eear

If new fermions are included, the anomaly cancellation conditions have
more solutions. We have found that including within each generation two fermions, 
$\psi^l$ and $\psi^e$, which under the SM
gauge group are vector-like and have the same charges as
$l_L$ and $e_R$, respectively, allows $z$ charges proportional to
$B-xL$ with $x$ arbitrary. The $U(1)_{B-xL}$ charges are shown in 
Table 1. This is the most general generation-independent 
charge assignment for the SM fermions that allows
quark and lepton masses from Yukawa couplings, and is relevant 
for $Z^\prime$ searches at the Tevatron. 


\begin{table}[t]
\centering
\renewcommand{\arraystretch}{1.5}
\begin{tabular}{|c||ccc||c|c|c|c|}\hline
& $SU(3)_C$ & $SU(2)_W$ & $U(1)_Y$ & $U(1)_{B-xL}$ & $U(1)_{q+xu}$ & $U(1)_{10+x\bar{5}}$ 
& $U(1)_{d-xu}$ \\ \hline\hline
$q_L$      &  3 & 2 & $ 1/3$   &   $1/3$    & $1/3$      &  $1/3$      & 0         \\
$u_R$      &  3 & 1 & $ 4/3$   &   $1/3$    & $x/3$      &  $-1/3$     & $-x/3$    \\
$d_R$      &  3 & 1 & $-2/3$   &   $1/3$    & $(2-x)/3$  &  $-x/3$     & $1/3$     \\ 
$l_L$      &  1 & 2 & $  -1$   &   $-x$     & $-1$       &   $x/3$     & $(-1+x)/3$\\
$e_R$      &  1 & 1 & $  -2$   &   $-x$     & $-(2+x)/3$ &  $-1/3$     & $x/3$     \\ \hline \hline
$\nu_R$    &    &   &          &   $-1$     & $(-4+x)/3$ &  $(-2+x)/3$ & $-x/3$    \\ [-1em]
           &  1 & 1 & 0        &            &            &             &           \\ [-1em]
$\nu_R^{\, \prime}$  &  &  &   &   $\cdot$  & $\cdot$    &    $-1-x/3$ &  $\cdot$  \\ \hline
$\psi_L^l$ &    &   &          &   $-1$     & $\cdot$    &  $-(1+x)/3$ & $-2x/5$   \\ [-1em]  
           &  1 & 2 & $  -1$   &            &            &             &           \\ [-1em]
$\psi_R^l$ &    &   &          &   $-x$     & $\cdot$    &  $2/3$      & $(-1+x/5)/3 $ \\ \hline
$\psi_L^e$ &    &   &          &   $-1$     & $\cdot$    &  $\cdot$    &  $\cdot$  \\ [-1em]  
           &  1 & 1 & $  -2$   &            &            &             &           \\ [-1em]
$\psi_R^e$ &    &   &          &   $-x$     & $\cdot$    &  $\cdot$    &  $\cdot$  \\ \hline
$\psi_L^d$ &    &   &          &  $\cdot$   &  $\cdot$   &  $-2/3$     & $(1-4x/5)/3$ \\ [-1em]  
           &  3 & 1 & $  -2/3$ &            &            &             &           \\ [-1em]
$\psi_R^d$ &    &   &          &  $\cdot$   & $\cdot$    &  $(1+x)/3$  & $ x/15 $ \\ \hline
\end{tabular}

\medskip
\parbox{5.5in}{
\caption{ Fermion gauge charges. 
\label{TableCharge}}}
\end{table}

All other generation-independent $U(1)_z$ charge assignments 
require the restrictions on fermion charges 
from fermion mass generation to be lifted, for example by replacing 
the Yukawa couplings with higher-dimensional operators. 
The six anomalies given in Eqs.~(\ref{eq:a33z})-(\ref{eq:aggz}) vanish
only for the nonexotic family of $U(1)_z$ charges 
that depends on two parameters \cite{Appelquist:2002mw}. 
Assuming that $z_q \neq 0$, and normalizing the 
$g_z$ gauge coupling such that $z_q = 1/3$, determines all other charges
as a linear function of a single free parameter, $x$, as shown in Table 1.
We label this charge assignment by $U(1)_{q+xu}$. Particular cases 
of $Z^\prime$ ``models'' include $U(1)_{B-L}$ for $x=1$,
the $U(1)_\chi$ from $SO(10)$ grand unification for   $x=-1$, and
the $[U(1)_R\times U(1)_{B-L}]/U(1)_Y$ group from left-right symmetric models
for $x=4 - 3g_R^2/g_Y^2$ where $g_R$ is the $U(1)_R$ gauge coupling.

Many popular $Z^\prime$ models are accessible at the Tevatron provided  
both the restrictions from fermion mass generation are lifted
and new fermions charged under the  SM 
gauge group are present.
We have found a couple of generation-independent charge assignments
that depend on a free parameter $x$, which include the $E_6$-inspired
$Z^\prime$ models that have been frequently analyzed by experimental
collaborations. Both require within each generation two fermions, 
$\psi^l$ and $\psi^d$, which under the SM
gauge group are vector-like and have the same charges as
$l_L$ and $d_R$, respectively. 
One assignment, labeled by $U(1)_{d-xu}$, has $z_q = 0$,  $z_d = 1/3$ and $z_u = -x/3$. 
For $x= 0$ the $E_6$-inspired $U(1)_I$ is recovered,
while $x=1$ gives the ``right-handed'' $U(1)_R$ group.
The other  assignment is such that all fermions  belonging to the 
$10$ representation of the $SU(5)$ grand unified group have the same 
$U(1)_z$ charge, assumed to be nonzero and normalized to $1/3$,
while the fermions  belonging to the $\bar{5}$ representation
have charge $x/3$. We label this  assignment by $U(1)_{10+x\bar{5}}$.
Anomaly cancellation requires two right-handed neutrinos 
per generation. 
Particular $E_6$-inspired
cases include $U(1)_\psi$ for $x=1$, 
$U(1)_\chi$  for $x=-3$, and $U(1)_\eta$  for $x=-1/2$.
Note that when the LEP and Tevatron experimental collaborations 
refer to these particular models, the gauge coupling is usually 
assumed to be determined by a unification relation: $g_z^2 = g_Y^2 \xi$,
with $\xi= 5/8$ for $U(1)_\psi$, $\xi= 3/8$ for $U(1)_\chi$,
$\xi= 1$ for $U(1)_\eta$, $\xi= 5/3$ for $U(1)_I$.
Thus, our families of models completely describe 
the physics of the GUT-inspired $U(1)$'s,
but also allow one to relax their assumptions and explore more 
general $Z^\prime$ physics.

There is an interesting class of $Z^\prime$ models in which the Higgs is a
pseudo-Goldstone boson of a spontaneously broken global symmetry.  These
``Little Higgs'' models always include at least one $Z^\prime$ to cancel the
leading quadratic divergence in the Higgs mass from loops of the ordinary
$W$ and $Z$.  A proto-typical model of this type is the ``Littlest Higgs''
\cite{Arkani-Hamed:2002qy}.  
This already reveals a key feature of the little Higgs 
$Z^\prime$: it always couples to the Higgs, and thus generically has strong
constraints from $Z$-$Z^\prime$ mixing,
requiring the $Z^\prime$ mass to be larger than several hundred GeV 
\cite{Csaki:2002qg}.  Thus, it is usually not very interesting 
for Tevatron searches.

\subsection{LEP II limits on $Z^\prime$ models}
\label{subsec:lep}

The constraints from  $e^+ e^-$ colliders on the $Z^\prime$ properties
fall into two categories: precision measurements
at the $Z$ pole, through the $Z$-$Z^\prime$ mixing discussed previously,  
and  measurements of $e^+ e^- \rightarrow f \bar{f}$ 
above the $Z$-pole at LEP-II, 
where $f$ are various SM fermions.
In practice, the agreement with
the SM requires that either the $Z^\prime$ gauge 
coupling is smaller than or of order $10^{-2}$ \cite{Appelquist:2002mw},
or else $M_{Z^\prime}$ is larger than the largest collider 
energy of LEP-II, of about 209 GeV. In the latter case, of 
interest at the Tevatron,
one can perform an expansion in
$s / M^2_{Z^\prime}$, where $s$ is the square of the center-of-mass energy.
This leads to effective contact interactions 
which have been bounded by LEP-II for all possible chiral structures
and for various combinations of fermions.
These interactions are parameterized by the LEP electroweak working group
\cite{lep:2003ih} as,
\begin{eqnarray}
\label{eq:ctlep}
\frac{\pm 4 \pi}{(1 + \delta_{ef}) (\Lambda^{f\pm}_{AB})^2} 
\left( \bar{e} \gamma_{\mu} P_A e \right)
\left( \bar{f} \gamma^{\mu} P_B f \right)
\end{eqnarray}
where $P_A$ is a projector for right- ($A=R$) or left-handed ($A=L$)
chiral fields,
and $\delta_{ef} = 1,0$ for $f=e$, $f \not = e$, respectively.  
These contact interactions 
provide a model-independent framework in which LEP-II data can confront
high mass effects beyond the SM, up to 
corrections of order $s / M^2_{Z^\prime}$.

In the absence of flavor violation in the $Z^\prime$ couplings, the $Z^\prime$
contributions to $e^+ e^- \rightarrow f \bar{f}$ for $f \not = e$ proceed through
an $s$-channel $Z^\prime$ exchange, with tree level matrix element,
\bear
\label{eq:ctzp}
\frac{g_z^2}{M_{Z^\prime}^2 - s}
\left[ \bar{e} \gamma_{\mu} (z_l P_L + z_e P_R) e \right]
\left[ \bar{f} \gamma^{\mu} (z_{f_L} P_L + z_{f_R} P_R) f \right]
\eear
where tiny terms of order $m_f m_e / M_{Z^\prime}^2$ have been dropped.
In the case of $f=e$, there is also a $t$-channel exchange, which in fact
motivates the factor of $(1+\delta_{ef})$ in Eq.~(\ref{eq:ctlep}), which
allows one to treat all $f$ equivalently at the level of matrix elements.

One should compare the LEP $\Lambda_{LL}$, 
$\Lambda_{RR}$, $\Lambda_{LR}$, and $\Lambda_{RL}$ limits to the
operators of each structure in the $Z^\prime$ theory in order to find
a limit on a given $Z^\prime$ model.  This procedure
finds the best {\em single} bound from each operator on a given
$Z^\prime$ model, but it ignores the potentially stronger bound that comes
from the combined effect of more than one operator.
In the absence of a dedicated reanalysis
of the data, this is the best one may do.
However, we reiterate that it does not always represent
the best potential bound from the data,
due to correlated effects on observables which cannot
be taken into account correctly in this way.
Typically the strongest bound comes from a single choice of
chiral interaction combination and $f$.

Matching the $Z^\prime$ matrix elements, Eq.~(\ref{eq:ctzp}), 
to the LEP-II formalism, Eq.~(\ref{eq:ctlep}),
one derives bounds such as,
\bear
{M_{Z^\prime}^2 - s} \geq \, \frac{g_z^2}
{4 \pi} |z_{e_A} z_{f_B}|  \, (\Lambda_{AB}^{f \pm})^2 ~,
\eear
and one must choose $\Lambda^+$ for $z_{e_A} z_{f_B} > 0$ or $\Lambda^-$
for $z_{e_A} z_{f_B} < 0$.  Typically, the LEP-II bounds on the $\Lambda$'s
are on the order of 10 TeV, schematically 
translating into bounds on the $Z^\prime$ mass
on the order of $M_{Z^\prime} \gae |z| g_z \times ({\rm a~few~TeV})$.
More precise bounds in the case of certain models are discussed 
below. 
The Tevatron can
effectively improve our knowledge of $Z^\prime$ models
only when the couplings $z g_z$ are appropriately small such that the LEP-II
limits are evaded for $Z^\prime$ masses in the several hundred GeV
range, but large enough that the $Z^\prime$ rate is observable compared
to backgrounds.

\begin{figure}[t]
\begin{center}
\includegraphics[width=12cm]{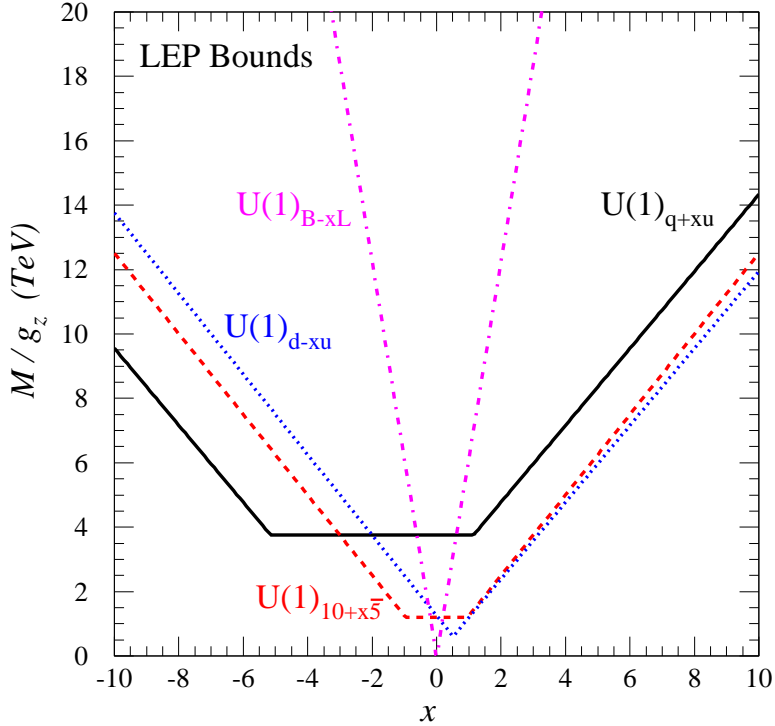}
\end{center}
\caption{Lower bounds on $M_{Z^\prime} / g_z$ from the LEP-II search for
$LL$, $RR$, $LR$ and $RL$ contact interactions, applied to the models of 
Table~\protect{\ref{TableCharge}} as a function of the continuous 
parameter $x$.  For $U(1)_{B-xL}$, we have included the bound on vector-like
$e^+ e^- \rightarrow \ell^+ \ell^-$, as is appropriate for that model.
}\label{fig:lepbound}
\end{figure}

For example, $U(1)_{B-xL}$ has vector-like interactions with quarks
and leptons, and thus it is better to compare LEP limits for vector-like
interactions ($\Lambda_{VV}$) than for the individual chiral set 
of $\Lambda_{LL}$, 
$\Lambda_{RR}$, $\Lambda_{LR}$, and $\Lambda_{RL}$.
From Ref.~\cite{lep:2003ih}, we see that the strongest of the 
$\Lambda^+_{VV}$ bounds is from the process
$e^+ e^- \rightarrow \ell^+ \ell^-$: $\Lambda^+ > 21.7$ TeV.
This is in fact one of
the most stringent bounds to be found from LEP-II.
Translating this bound in the specific
couplings of $U(1)_{B-xL}$ results in the limit
$M_{Z^\prime} \geq |x| g_z \times ({\rm 6~TeV})$.
For the remaining models of Table~\ref{TableCharge}, 
the analysis is more
complicated, because the best bounded channel 
typically depends on the value of $x$.
Thus, for these models, we have scanned (for fixed $-10 \leq x \leq 10$) 
through all channels 
of $\Lambda_{LL}$, $\Lambda_{RR}$, $\Lambda_{LR}$, $\Lambda_{RL}$ 
and chosen the best bound.  The results are shown in Figure~\ref{fig:lepbound}.
It is interesting to note that for $|x| \gae 1$, $U(1)_{B-xL}$ is very
strongly constrained by LEP II data, whereas for $x \rightarrow 0$ the 
coupling to electrons becomes small and the bounds disappear.

We have compared our results for general $x$ with the $E_6$-inspired
models studied in detail at LEP-II \cite{lep:2003ih}. 
As explained above, these correspond to
specific points in $x$ for a given model family.  We find that our results
agree with the LEP bounds for the dedicated analysis at or better than 
the $25\%$ level (depending on the model), 
thus indicating that our procedure does a good job in
comparison with the dedicated analysis for the points
in which the two may be compared.  For most of the parameter 
space there is no dedicated LEP analysis, and we present for the first time
the LEP bounds on the general class of $Z^\prime$ models.

Generically, the fact that LEP II was an $e^+ e^-$ collider implies
that these strong bounds can be evaded by a $Z^\prime$ which couples only
very weakly to electrons\footnote{For example, theories such as
Top-color assisted Technicolor \cite{Hill:1994hp},
Top-flavor \cite{Chivukula:1995gu}, 
or Supersymmetric Top-flavor \cite{Batra:2003nj}, have
small couplings to the first generation.}.  
Also, different chiral structures than the 
vector-like interactions of $U(1)_{B-xL}$ can have weaker bounds.  From
Ref.~\cite{lep:2003ih} one observes that a $Z^\prime$ which couples only to
left-handed or right-handed electrons is bounded only by
$\Lambda^+ \geq 7.1, 7.0$ 
TeV (for $\Lambda_{LL}$ and $\Lambda_{RR}$, respectively).
This implies a bound of
$M_{Z^\prime} \geq g_z z \times ({\rm 1.9~TeV})$, approximately three times
weaker than the bound on $U(1)_{B-xL}$.

\section{$Z^\prime$ Searches at the Tevatron}
\label{sec:tevatron}
\setcounter{equation}{0}

At the Tevatron, searches for additional neutral gauge bosons can be performed 
in a variety of processes. If such bosons couple to the SM quarks, they
may be directly observed through their production and subsequent decay 
into high energy lepton pairs or jets. The case of the decay into leptons
is particularly attractive due to low backgrounds and good momentum 
resolution.  Bounds on several models containing extra neutral gauge bosons, 
have been set by both the CDF \cite{Abe:1991vn,Abe:1994ns,Abe:1997fd} 
and D0 \cite{Abachi:1996ud,Abbott:1998rr,Abazov:2001qd} experiments by  
measuring high energy lepton pair production cross sections. 
Searches have been made in the $e^+e^-$ channel, which has the best
acceptance, and thus best systematics, as well as in the $\mu^+\mu^-$ channel.
More recently, the challenging $\tau^{+}\tau^{-}$ channel has also been 
analyzed \cite{FNALJET}. 
The $\mu^+\mu^-$ and $\tau^{+}\tau^{-}$ final states, along with 
the $Z^\prime$ decay into jets which suffers from huge QCD backgrounds, 
can probe $Z^\prime$ bosons with suppressed couplings to the electrons, 
which are not constrained by the LEP searches.

In what follows, we will restrict ourselves to the 
study of the leptonic decay modes, proposing a simple, model-independent, 
parameterization for the $Z^{\prime}$ production cross section and analyzing
its theoretical and experimental feasibilities and limitations.

\subsection{$Z^{\prime}$ Hadro-production}

The additional terms, beyond those coming from SM particles, 
in the differential 
cross section for production of a pair of charged leptons due to the
presence of an extra neutral gauge boson can be written as 
\cite{Hamberg:1990np}
\begin{equation}\label{eq:cs1}
\frac{d}{dQ^2}\sigma \left( p\bar{p} \rightarrow Z^\prime X \rightarrow l^+l^- X \right)
=\frac{1}{s}\,\sigma(Z^\prime\rightarrow
l^+l^-)\,W_{Z^{\prime}}\left(s,Q^2\right)
+\frac{d\sigma_{\mbox{\small int}}}{dQ^2}\,,
\end{equation}
where $Q$ is the invariant mass of the lepton pair, and $\sqrt{s}$ 
is the energy of the $p\bar{p}$ collision in the center-of-momentum
frame. The first term accounts for the contributions from
the $Z^\prime$ itself and has been explicitly factorized into a 
hadronic structure function, $W_{Z^{\prime}}$, 
containing all the QCD dependence and the couplings of quarks 
to the $Z^\prime$, and 
\begin{equation}\label{eq:slep}
\sigma(Z^\prime\rightarrow l^+l^-)
=\frac{g_z^2}{4\pi}\,
\left( \frac{ z_{l_j}^2+z_{e_j}^2 }{288} \right) \,
\frac{Q^2}{(Q^2-M_{Z^{\prime}}^2)^2+M_{Z^{\prime}}^2\,
\Gamma_{Z^{\prime}}^2}\,.
\end{equation} 
Up to NLO in QCD, only the partonic processes 
$\bar{q}{q}\rightarrow Z^\prime X$ 
(non-singlet) and $qg\rightarrow Z^{\prime}X$ contribute to the hadronic 
structure function. 
If the $Z^{\prime}$ couplings to quarks are generation independent, 
both processes give contributions which are proportional
to $(z_{q}^2+z_{u}^2)$ or $(z_{q}^2+z_{d}^2)$ for up and down type quarks, respectively.
Therefore, the hadronic structure function can be written as  
\begin{equation}\label{eq:hsf}
\begin{split}
W_{Z^{\prime}}\left(s,M_{Z^\prime}^2\right)=
g_z^2\,\Bigg[(z_q^2+z_u^2)\,
w_{u}\left(s,M_{Z^\prime}^2\right)
+
(z_q^2+z_d^2)\,
w_{d}\left(s,M_{Z^\prime}^2\right)
\Bigg]\, .
\end{split}
\end{equation}
The functions $w_u$ and $w_d$ 
do not depend on any coupling and are exactly
the same for any model containing neutral gauge bosons coupled in a 
generation 
independent way to quarks. In the $\overline{\mbox{MS}}$ scheme, they 
are given by
\bear\label{eq:wuwd}
\begin{split}
w_{u(d)}=\sum_{q=u,c(d,s,b)}&\,\int_0^1 dx_1\,\int_0^1 dx_2 \,\int_0^1 dz
\,\Big\{f_{q/P}(x_1,M_{Z^\prime}^2)
\,f_{\bar{q}/\bar{P}}(x_2,M_{Z^\prime}^2)
\,\Delta_{qq}(z,M_{Z^\prime}^2)\\ \vspace*{0.8em}
&+f_{g/P}(x_1,M_{Z^\prime}^2)\, 
\left[ f_{q/\bar{P}}(x_2,M_{Z^\prime}^2) +
f_{\bar{q}/\bar{P}}(x_2,M_{Z^\prime}^2) \right]\,
\Delta_{gq}(z,M_{Z^\prime}^2)\\ \vspace*{0.8em} 
&+(x_1\leftrightarrow x_2,P\leftrightarrow \bar{P})
\Big\}\,\delta\left(
\frac{M_{Z^\prime}^2}{s} - z\,x_1\,x_2\right)
\,,
\end{split}
\eear
where $f_{i/P}(x,M^2)$ and $f_{i/\bar{P}}(x,M^2)$ are the PDFs 
for the proton and antiproton, respectively, and
\begin{eqnarray}
&&\begin{split}
\Delta_{qq}(z,M_{Z^\prime}^2)&=\delta(1-z)+\frac{\alpha_s(M_{Z^\prime}^2)}
{\pi}\,C_F\,\Bigg[ \delta(1-z) \left(\frac{\pi^2}{3} - 4\right)
+4\left(\frac{\ln(1-z)}{1-z}\right)_{\! +z[0,\underline{1}]}\\
&-2\,(1+z)\,\ln(1-z)-\frac{1+z^2}{1-z}\,\ln z
\Bigg]\,,
\end{split}\\ [0.5em]
&&\begin{split}
\Delta_{gq}(z,M_{Z^\prime}^2)&=\frac{\alpha_s(M_{Z^\prime}^2)}
{2\pi}\,T_F\,\Bigg[\left(1-2z+2z^2\right)\,\ln\frac{(1-z)^2}{z}
+\frac{1}{2}+3z-\frac{7}{2} z^2
\Bigg]\,.
\end{split}
\end{eqnarray}
The color factors are $C_F=4/3$ and $T_F=1/2$, and we have set the 
renormalization and factorization scales to $M_{Z^\prime}$. 

The second term in Eq. (\ref{eq:cs1}), $d\sigma_{\mbox{\small int}}/dQ^2$,
corresponds to the interference of the $Z^\prime$ with
the $Z$ and the photon. If the $Z^\prime$ resonance is narrow enough, 
the interference of the $Z^\prime$ with
the $Z$ and photons can be neglected (see the Appendix). 

In the narrow width approximation, the expression for the total cross 
section is simply obtained from the differential cross section, explicitly
\begin{equation}\label{eq:tcs1}
\sigma\left( p\bar{p} \rightarrow Z^\prime X \rightarrow l^+l^- X \right)
=\frac{\pi}{48\,s}\,W_{Z^{\prime}}
\left(s,M_{Z^\prime}^2\right)
\,\mbox{Br}(Z^{\prime}\rightarrow l^{+}l^{-})\,,
\end{equation}
where $\mbox{Br}(Z^{\prime}\rightarrow l^{+}l^{-})$ is the branching ratio for
the decay of $Z^\prime$ into the corresponding pair of leptons.
Using the expression of the hadronic structure function, Eq.~(\ref{eq:hsf}),
one obtains
\begin{equation}\label{eq:cs2}
\sigma\left( p\bar{p} \rightarrow Z^\prime X \rightarrow l^+l^- X \right)
=\frac{\pi}{48\,s}\,\left[c_{u}
\,w_{u}\left(s,M_{Z^\prime}^2\right)+
c_{d}
\,w_{d}\left(s,M_{Z^\prime}^2\right)
\right]\,.
\end{equation}
The coefficients $c_u$ and $c_d$, given by
\begin{equation}
c_{u,d}=g_z^2\,(z_q^2+z_{u,d}^2)\,
\mbox{Br}(Z^{\prime}\rightarrow l^{+}l^{-})\,,
\end{equation}
contain all the dependence on the couplings of quarks and leptons to 
the $Z^\prime$, while $w_u$ and $w_d$ only depend on the mass of the
gauge boson and can be calculated in a completely model-independent way.

The parameterization given in Eq.~(\ref{eq:cs2}) permits a
direct extraction of a bound in the $c_u-c_d$ plane from the 
experimental limit 
for the cross section, which can be later 
compared to the predictions of particular models. This fact is particularly
useful for models admitting free parameters like the ones discussed in the
preceding sections. In particular, these quantities are simply computed for 
a given $Z^\prime$ model, without need to compute the hadro-production cross
section, and thus are a common ground between theory and experiment. 

\begin{figure}[t]
\begin{center}
\includegraphics[width=12cm]{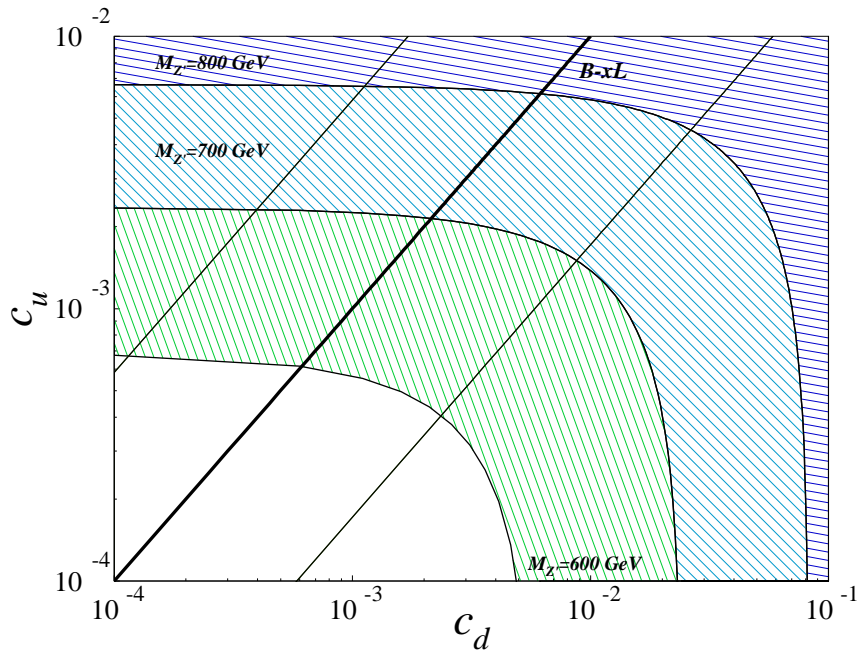}
\end{center}
\caption{Excluded regions in the $c_d-c_u$ plane 
from the current 95\% C.L. limit for $\sigma\cdot\mbox{Br}(Z^\prime\rightarrow l^{+}l^{-})$ 
given in \cite{CFD:CL},
for different values of the 
$Z^\prime$ mass. The thick straight line corresponds to values of $c_u$ and $c_d$ 
in the $B-xL$ model, in which $c_u=c_d$. The area between the two thin 
straight lines is the region where the $q+x u$ model lies.
}\label{fig:exclcucd}
\end{figure}

The D0 and CDF Collaborations have set preliminary 95\% C.L. limits
for $\sigma\cdot\mbox{Br}(Z^\prime\rightarrow l^{+}l^{-})$ in 
Run II with 200 fb$^{-1}$
\cite{D0:CL, CFD:CL}. In 
Figure \ref{fig:exclcucd} we show the excluded regions in the $c_d-c_u$
plane for different values of the mass of the $Z^\prime$ boson as obtained from
the limit on 
$\sigma\cdot\mbox{Br}(Z^\prime\rightarrow l^{+}l^{-})$ given 
by the CDF Collaboration \cite{CFD:CL} in Run II with 200 fb$^{-1}$. 
Very similar results are obtained using the results by the D0 Collaboration
\cite{D0:CL}.
The $w_u$ and $w_d$ coefficients in Eq. (\ref{eq:cs2}) for this plot were 
calculated at NLO with MRST02 PDFs \cite{Martin:2002dr}.  From the current
generation of CTEQ \cite{Pumplin:2002vw} 
PDFs, one obtains very similar results. 

It is instructive to compare these limits with the predictions of the four families 
of models presented in Section 2. The values of $c_u$ and $c_d$ as functions of the 
gauge coupling $g_z$ and the $x$ parameter are given in Table 2
\begin{table}[t]
\centering
\renewcommand{\arraystretch}{2.5}
\begin{tabular}{|c||c|c|c|c|}\hline
& $U(1)_{B-xL}$ & $U(1)_{q+xu}$ & $U(1)_{10+x\bar{5}}$ & $U(1)_{d-xu}$ \\ \hline\hline
$c_u/g_z^2$ & {\large $\frac{4x^2}{9\left(4+9x^2\right)}$ } & 
{\large $\frac{\left(1+x^2\right)\left(13+4x+x^2\right)}{27\left(40-8x+7x^2\right)}$ }
& {\large $\frac{ 2\left( 1+x^2 \right) }{ 135\left(2+x^2\right) }$ }
& {\large $\frac{ x^2\left( 1-2x+2x^2 \right) }{ 27\left(5-4x+6x^2\right) }$ } \\ \hline
$c_d/c_u$ & $1$ & $1 + 4${\large $\frac{1-x}{1+x^2}$ }  
& {\large $\frac{1+x^2}{2}$ }  & {\large $\frac{1}{x^2}$}  \\ \hline
\end{tabular}

\medskip
\parbox{15cm}{
\caption{Predictions for $c_d$ and $c_u$ in four families of models defined in Table 1. The
branching fractions are computed at tree level for $M_Z^\prime > 2m_t$ and assuming decays 
only into SM particles.
\label{TablePred}}}
\end{table}
In Figure \ref{fig:exclcucd} displays 
the values of $(c_d,c_u)$ corresponding to the $B-xL$ and $q+x u$ models.
In the $B-xL$ case, these points are constrained to satisfy $c_u=c_d$, 
corresponding to the thick straight line. 
For the $q+x u$ model, the allowed region
is 
\bear
(3-2\sqrt{2})\,c_d\le c_u\le (3+2\sqrt{2})\,c_d ~,
\eear
which corresponds to
the area between the two thin straight lines in Figure \ref{fig:exclcucd}. 
The $10+x\,\bar{5}$ model, in turn,
is constrained to the region $c_u\le 2\,c_d$, whereas there are no constraints
for the possible values of $c_d$ and $c_u$ in the $d-x u$ model. 


\subsection{Higher-order corrections}

\begin{figure}[t]
\begin{center}
\includegraphics[width=12cm]{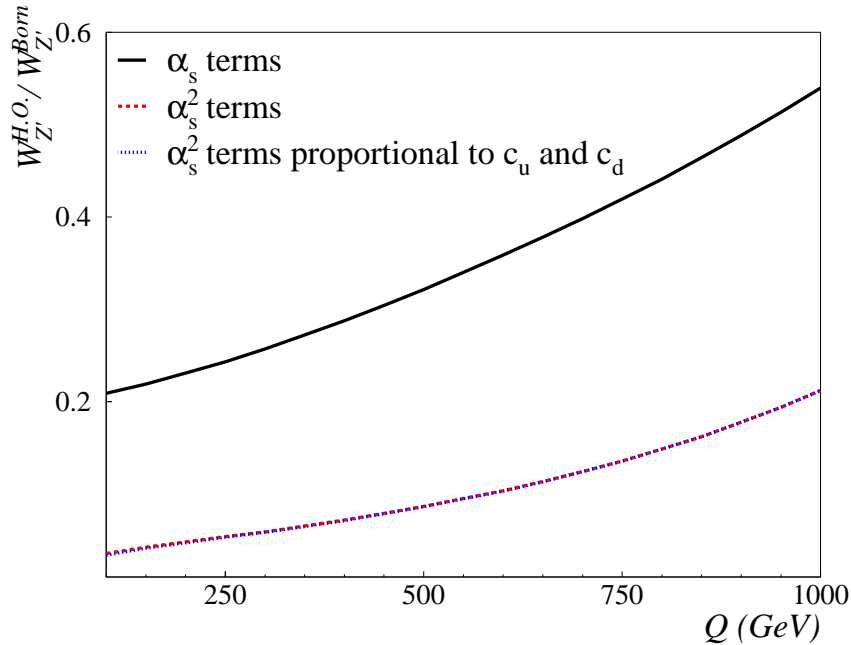}
\end{center}
\caption{The ratio of the ${\cal O}(\alpha_s)$ and ${\cal O}(\alpha_s^2)$ 
corrections to the hadronic structure function 
over the Born contribution, assuming SM couplings for the $Z^\prime$.
The full line corresponds to the
${\cal O}(\alpha_s)$ corrections. The dashed and dotted lines, which, within
the resolution of the figure, appear as a single line, correspond to 
the total ${\cal O}(\alpha_s^2)$ contributions and to
the ${\cal O}(\alpha_s^2)$ corrections retaining only those terms that 
give contributions proportional to $c_u$ and $c_d$ to the cross section,
respectively. 
}\label{fig:nnlocorr}
\end{figure}

At NNLO in QCD, the general expression in Eq. (\ref{eq:hsf}) is no longer 
valid. This is due to the contributions from the partonic processes 
$\bar{q}q\rightarrow Z^\prime X$ (singlet) and $qq\rightarrow Z^\prime X$, 
which depend upon a variety of coupling combinations in addition
to $(z_q^2+z_u^2)$ and $(z_q^2+z_d^2)$. Thus, one should worry whether 
$c_u$ and $c_d$ are a sufficient description of the model.
The actual size of these corrections can be estimated
by looking at a particular model. In Figure \ref{fig:nnlocorr} we plot the 
sizes of the ${\cal O}(\alpha_s)$ and ${\cal O}(\alpha_s^2)$ terms 
to the structure function at NNLO relative to the Born contribution for the
case of SM-like couplings of the $Z^\prime$ boson. The NNLO corrections were
calculated with the program {\sc ZPROD} 
\cite{Hamberg:1990np,Harlander:2002wh,vNeerven:zwprod}
and the MRST02 NNLO set of PDFs  \cite{Martin:2002dr}. We have split
the ${\cal O}(\alpha_s^2)$ corrections in two parts, one proportional
to $c_d$ and $c_u$, and the other depending upon other combinations of 
the couplings. 
Contributions proportional to $c_d$ and $c_u$, coming from 
${\cal O}(\alpha_s^2)$ corrections to processes already present at lower 
orders, are clearly the dominant ones, overcoming the remaining pieces by
more than an order of magnitude in the whole $Q$ range.
Typically the terms with mixed couplings contribute less
than one per mil to the structure function, while the total 
${\cal O}(\alpha_s^2)$ corrections amount between 2\% and 20\% [for comparison,
the ${\cal O}(\alpha_s)$ terms contribute between 20\% and 50\% of the
structure function].

Although the actual values of the different higher order corrections
will depend upon the model considered, it is reasonable to expect that terms
not contributing to the $c_d$-$c_u$ piece of the cross section will be
negligible at NNLO. In that case, the parameterization in Eq. (\ref{eq:hsf})
holds as a very good approximation, and experimental bounds can be set in
a model independent way, taking $c_d$, $c_u$ and $M_{Z^\prime}$ as the
only relevant parameters at NNLO accuracy.

\begin{figure}[t]
\begin{center}
\includegraphics[width=12cm]{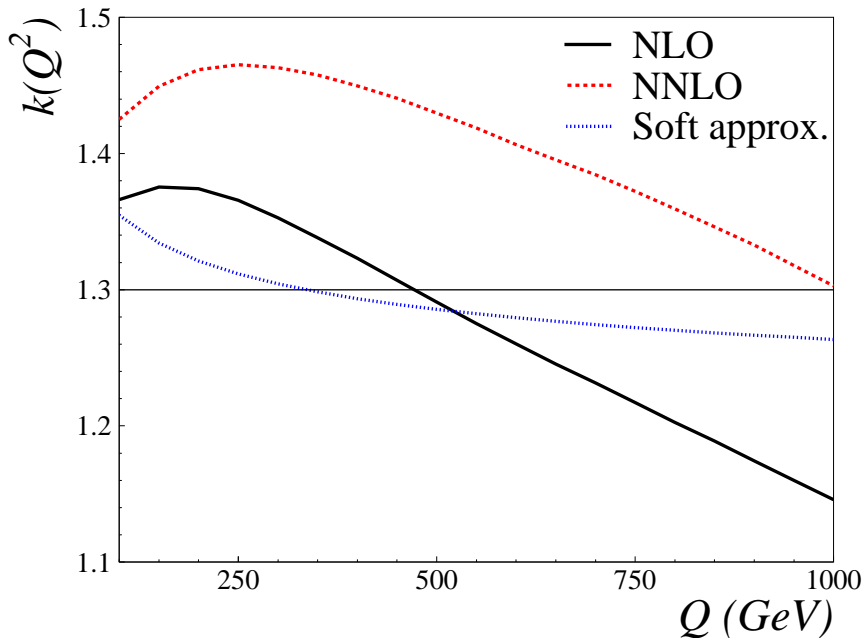}
\end{center}
\caption{
NLO and NNLO $k$-factors for SM-like couplings as a function of the 
invariant mass  of the lepton pair. Also shown is an approximation, Eq.~(\ref{eq:soft}), for the
NLO $k$-factor which only takes into account the soft corrections to the 
structure function in the DIS scheme.
}\label{fig:kfactor}
\end{figure}

It is customary to take into account higher order corrections to the structure
function by means of a $k$-factor, allowing the use of LO 
Montecarlo simulations, corrected afterwards by the mentioned factors. 
As the structure function at LO has a different dependence upon $Q$ than 
at NLO or NNLO, the corresponding $k$-factors, defined as the ratio of the 
higher order results over the LO one, 
$k_{N^iLO}=W_{Z^\prime}^{N^iLO}/W_{Z^\prime}^{LO}$, vary 
noticeably with the invariant mass
of the lepton pair. This variation has to be properly taken into account 
when correcting LO Montecarlo results. 

In Figure \ref{fig:kfactor} we plot the NLO and NNLO 
$k$-factors obtained again with the program {\sc ZPROD}, 
for SM-like couplings. 
There, we also show a usual approximation for the NLO $k$-factor (used, for instance, in 
Ref. \cite{Abe:1997fd}), which takes
into account only the soft pieces of the NLO structure function 
in the DIS scheme, namely
\bear\label{eq:soft}
k_{\mbox{soft}}=1+\frac{4}{3}\,\frac{\alpha_s(Q^2)}{2\,\pi}\,
\left(1+\frac{4}{3}\,\pi^2\right)~.
\eear
The variation of the $k$-factor, in the plotted range of $Q$, for 
the NLO case amounts to more than 10\%, the genuinely NNLO corrections are
also sizable whereas the soft approximation does not provide a good 
description of the higher order effects. The difference between the 
soft approximation and the full $k$-factors, yields a discrepancy 
of about 5\% to 10\% for the cross section
in the high mass region, which is particularly relevant in the search of 
extra gauge bosons. Similar results are obtained using a flat $k$-factor
with $k=1.3$, as in reference \cite{CFD:CL}, instead of the soft 
approximation.

So far we only discussed QCD corrections to the $Z^\prime$ cross section, 
which are of the order of 30\% as can be seen from the $k$-factors in Figure
\ref{fig:kfactor}. There are also corrections from the electroweak sector,
which we will address briefly. The complete ${\cal O}(\alpha)$ corrections to 
the SM contributions to neutral current Drell-Yan process were calculated in 
\cite{Baur:2001ze}. There, it was found that these corrections are large, 
particularly in the high invariant mass region, being of the order of 12\%
at Tevatron energies for $m_{ll}\simeq 700\mbox{ GeV}$ in the electron
channel. Besides affecting the background for the search of 
additional neutral gauge bosons, electroweak radiative corrections also
modify the signal cross section. However, as we will see, 
these effects are substantially smaller for the $Z^\prime$ terms.  

As shown in \cite{Baur:2001ze}, the main contributions to the electroweak 
corrections come from the box diagrams and cannot be factorized into 
effective couplings and masses. In particular, box diagrams with two
charged bosons give rise to large double logarithms which are the 
origin of the large corrections in the high mass region. On the other hand,
box diagrams that include neutral bosons ($\gamma$, $Z^{0}$ and $Z^\prime$)
always appear in combination with their crossed versions and that leads to a
cancellation of the double logs \cite{Ciafaloni:1998xg}. Then, the 
non-factorizable contributions that affect exclusively the signal cross 
section only include subleading simple logs and thus the corrections 
are expected to be smaller than for the SM background.

The remaining contributions come from QED and factorizable 
purely weak corrections. The later can always be absorbed into effective
couplings and masses, thus, they do not affect the signal cross section
where these quantities are treated as free parameters. The main
electromagnetic contributions 
come from large logarithms due to collinear
photon emission in the initial and final states, and 
affect both the signal and 
background cross sections. The large contributions coming from initial 
state radiation can be factorized
into the PDFs, modifying the DGLAP evolution equations for
the partonic densities. After factorization, the remaining terms 
are typically at the per mil level, reaching 1\% in the
high momentum fraction region \cite{Baur:2001ze} whereas the QED 
modifications to the evolution equations are small and neglected in 
comparison to the uncertainties in the PDFs \cite{Baur:1997wa}. 
However, collinear emission in the final state gives 
corrections of the order of 5\% for the electron channel in the 
high mass region, and a careful analysis should probably take them into
account. 

\subsection{Model dependence in experimental bounds}

As we have shown in the previous section, the parametrization given in Eq. 
(\ref{eq:cs2}) allows to extract model independent constraints on the
coefficients $c_d$ and $c_u$ from the experimental results for the
lepton pair production cross section. A key assumption for this analysis is
that the bounds for the cross section can be extracted from data in a model
independent manner. In particular, the experimental analyses involve 
corrections for the finite acceptance of the detectors to extract the total 
cross section. As the acceptance is obtained from detailed Monte Carlo 
simulations, which need to assume particular values of the couplings, it is 
far from trivial that this procedure does not introduce model dependence 
into the experimental bounds. 

In Ref. \cite{Abe:1991vn}, the changes in the acceptance with variations in the
couplings of up and down quarks were studied. There, it was found that the
acceptance changes very little when considering the limiting cases of either
decoupling up or down quarks. However, this study was limited to SM-like 
couplings for electrons, a feature that, a priori, might be too restrictive.  
To study the actual model dependence of the experimental acceptance, in this
section, we will consider the angular distribution of the lepton pair and apply
simple cuts on this distribution. For simplicity we will restrict to the LO 
approximation.

\begin{figure}[t]
\begin{center}
\begin{minipage}{7.5cm}
\begin{center}
\includegraphics[height=5.5cm]{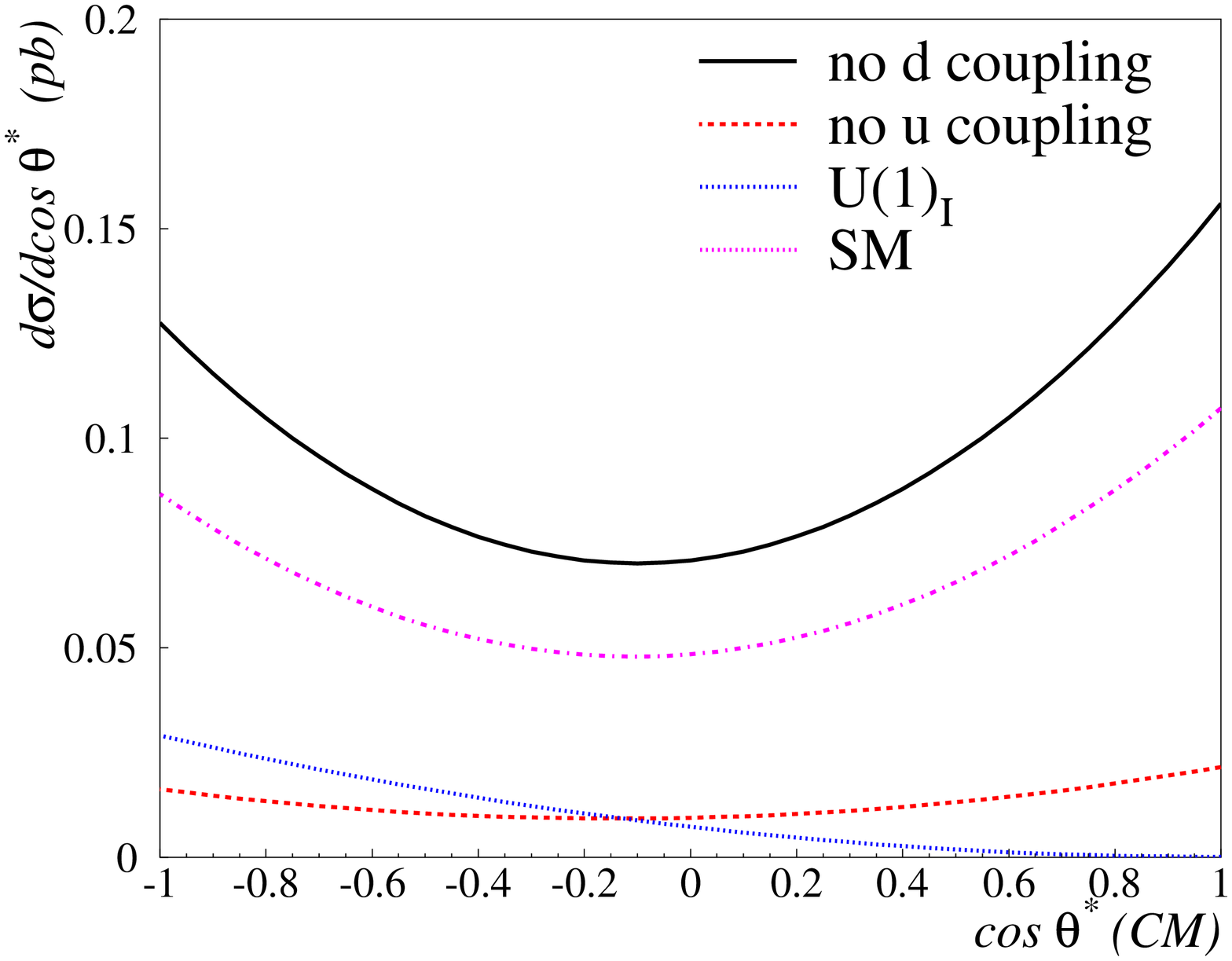}
{\bf (a)}
\end{center}
\end{minipage}
\begin{minipage}{7.5cm}
\begin{center}
\includegraphics[height=5.5cm]{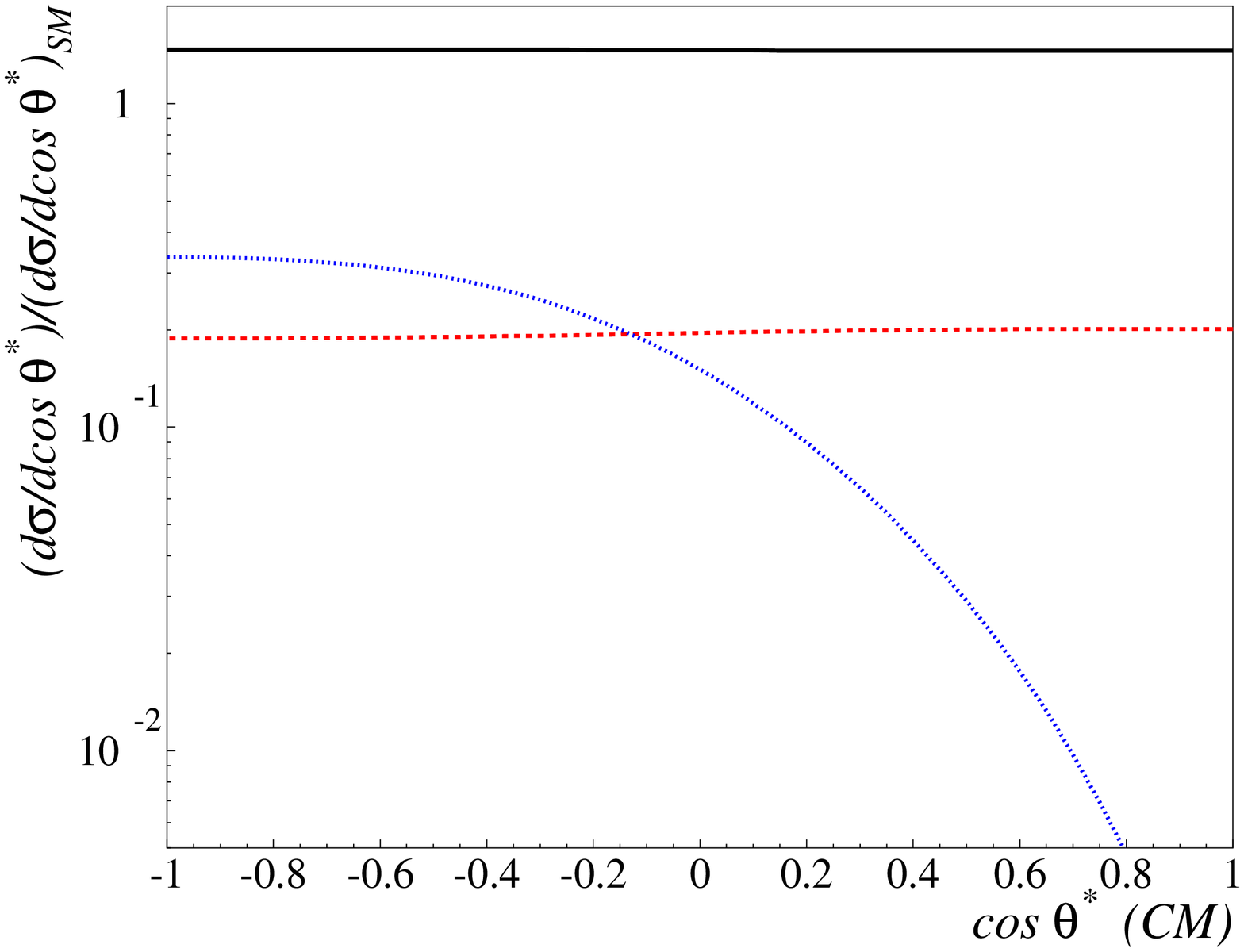}
{\bf (b)}
\end{center}
\end{minipage}
\end{center}
\caption{{\bf (a)} Angular distribution, in the CM frame of the lepton pair,
at LO in QCD for 
different models. The lines labeled no $u$ ($d$) coupling correspond to SM 
charges but neglecting the coupling of up (down) type quarks to the 
$Z^\prime$. 
{\bf (b)} Ratio between the differential cross section (shown in panel {\bf a)}
in different models and the case of SM-like charges.
}\label{fig:angdist}
\end{figure}

In the left panel of Figure \ref{fig:angdist} we plot the LO cross 
section differential with respect to the
azimuthal angle, in the center of mass frame of the lepton pair, 
with different assumptions for the couplings, setting 
$M_{Z^\prime}=600~\mbox{GeV}$. We considered SM-like couplings, SM-like
couplings with up or down couplings neglected and the $\mbox{E}_6$ 
inspired model $U(1)_I$ mentioned in Section 2. 
The right panel shows the ratios between 
the cross section in the last three cases to the cross 
section with SM couplings. Except for the overall normalization, the 
cases where the $Z^\prime$ does not couple to either up or down type quarks 
differ very little from the SM case. This feature can be traced back to the
peculiar fact that, in the SM, the left and right handed couplings of 
charged leptons, satisfy $z_l^2-z_e^2=\sin^2\theta_W-1/4
\simeq 0.02$. Then, the terms odd under $\theta\rightarrow -\theta$ 
are suppressed relative to the even ones in the LO differential cross 
section, which turns out to be 
nearly symmetric. This characteristic feature is in sharp contrast with 
the behavior in the
$U(1)_I$ case, where the asymmetry is almost maximal. This is related to the  
vanishing of the couplings of the right-handed electrons and left-handed
down-type quarks to the $Z^\prime$ in this last model.

\begin{figure}[ht]
\begin{center}
\includegraphics[height=7.5cm]{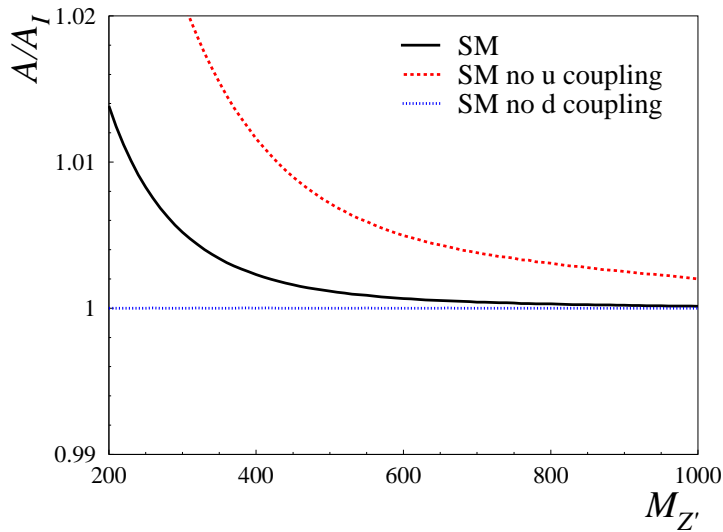}
\end{center}
\caption{The Acceptance ($A$), computed
from the LO cross section integrated over the azimuthal angle of the 
leptons in the laboratory frame in the region 
$50^o\le\theta_{\mbox{lab}}\le 130^o$. We 
considered the cases of SM couplings, SM couplings without $u$-quark couplings
and without $d$-quark couplings respectively. 
To show the variation compared to a most extreme
situation, we normalize to the corresponding LO acceptance of 
the $U(1)_I$ model. 
}\label{fig:acceptance}
\end{figure}

To get a handle on how the noticeable differences in the angular distribution
affect the experimental acceptance, we crudely estimated it 
by integrating the differential cross section imposing a symmetric
cut on the lepton angle, defined in the laboratory frame, and normalizing
to the total cross section. In Figure \ref{fig:acceptance} we plot the results 
obtained for the models considered in the previous paragraph, taking  
the $U(1)_I$ case as reference normalization.
Although the angular distribution in this model differs substantially
from that in the the other cases, the corresponding acceptances 
coincide at the few percent level. Note that the SM case without down-type quark couplings 
is practically indistinguishable from the $U(1)_I$ one. 

This, apparently peculiar, result for the acceptance is due to the fact that,
on average, the vector boson is produced with very small longitudinal momentum,
so the boost to go from the laboratory frame to the center of mass one is
small. Thus, a symmetric cut in the lab frame corresponds to an almost 
symmetric cut in the center of mass frame, which in turn means that the
ratio we are using to estimate the acceptance has only small contributions
that depend on the couplings of the quarks and leptons to the $Z^\prime$.  In 
particular, models that do not couple at all to up (down) type quarks give 
practically identical results for the acceptance. 

In conclusion, we find that, even though the angular distribution of the
final state leptons is highly model dependent,
the experimental acceptance, which we naively estimate through angular cuts,
is only mildly affected by this dependence, in accordance with previous 
studies by the experimental groups. On the other hand, the model 
dependence of the
angular distribution potentially can be a tool to discriminate between models
in case of a discovery. 

\begin{figure}[h!]
\begin{center}
\includegraphics[width=7.5cm]{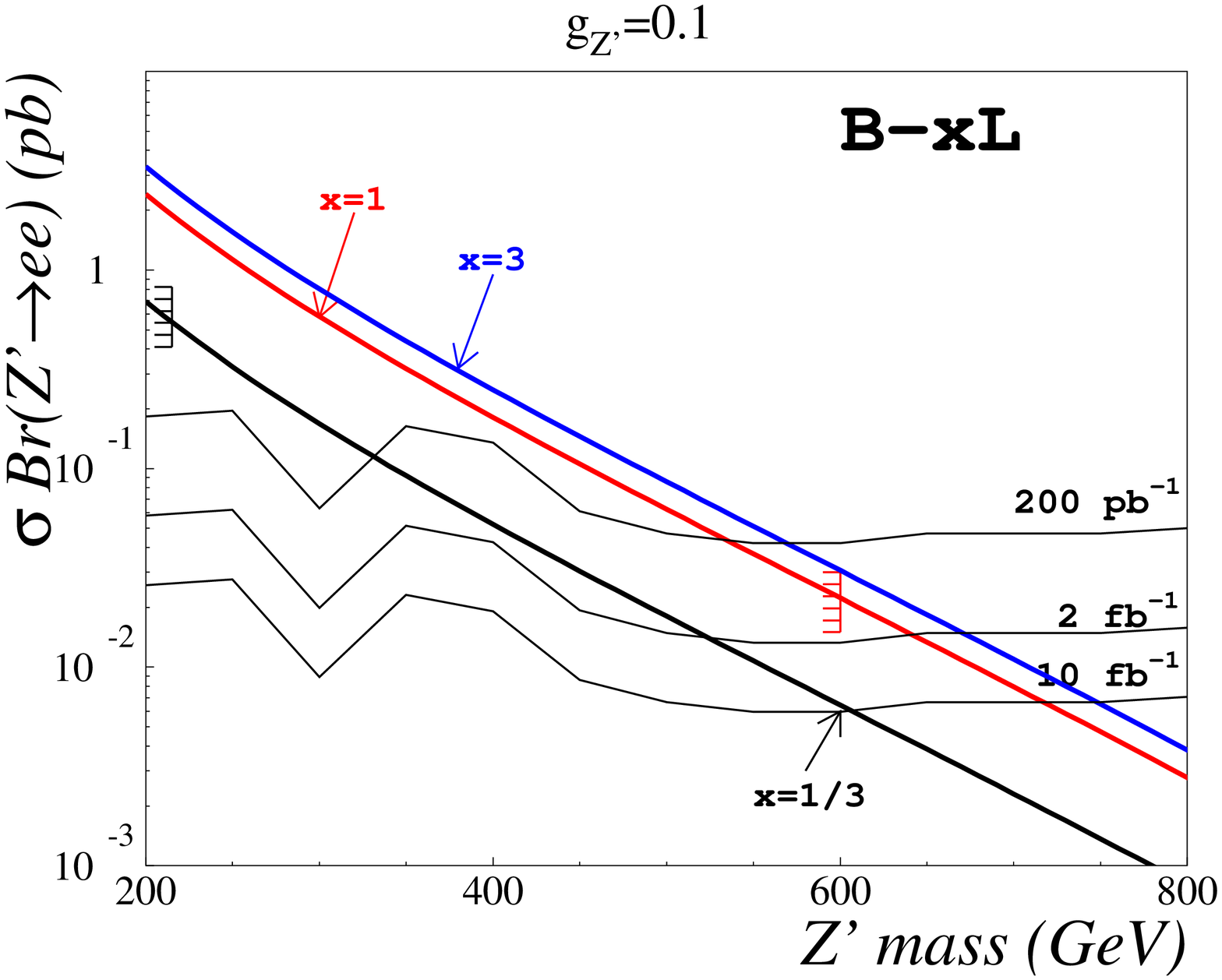}
\includegraphics[width=7.5cm]{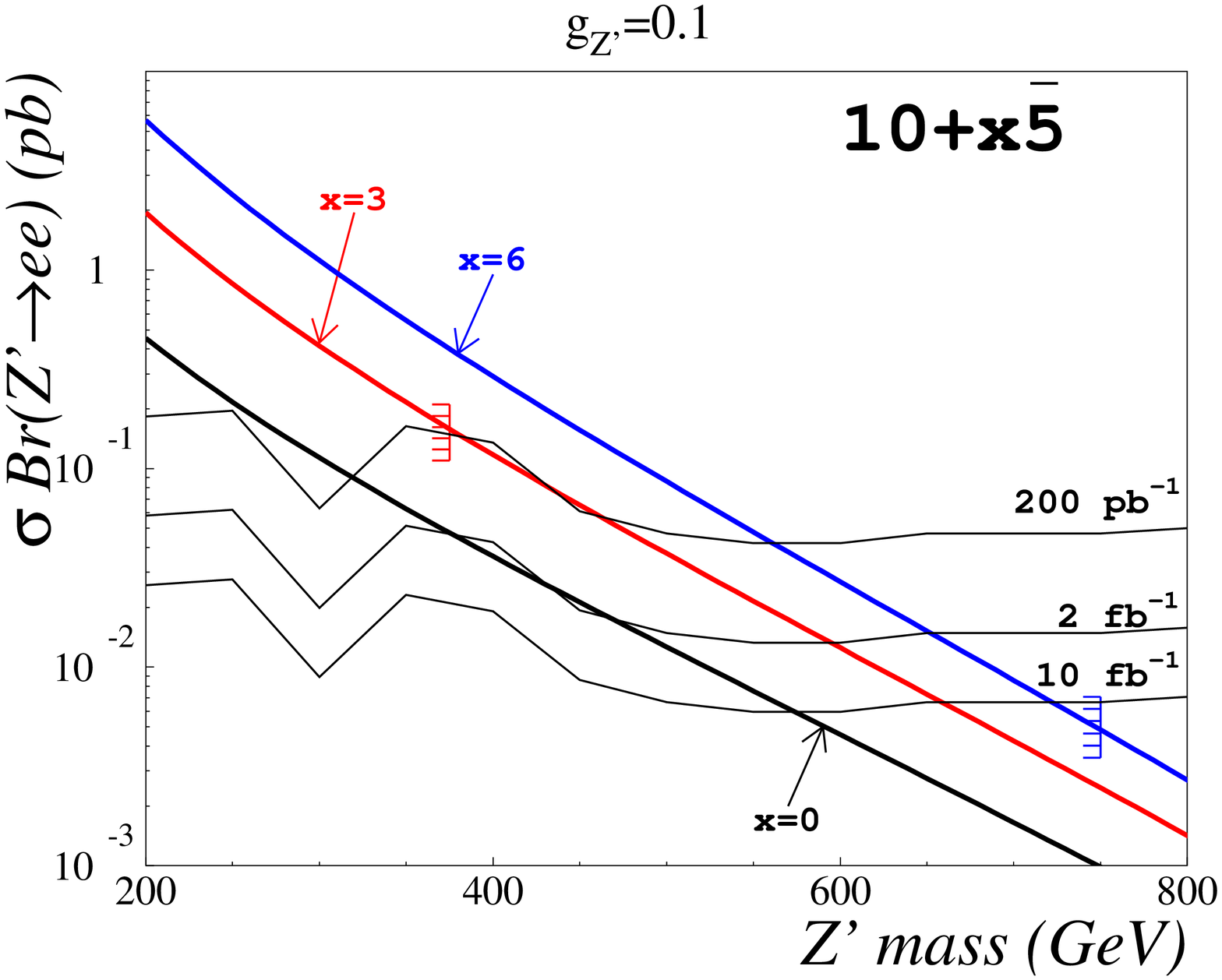}
\includegraphics[width=14cm]{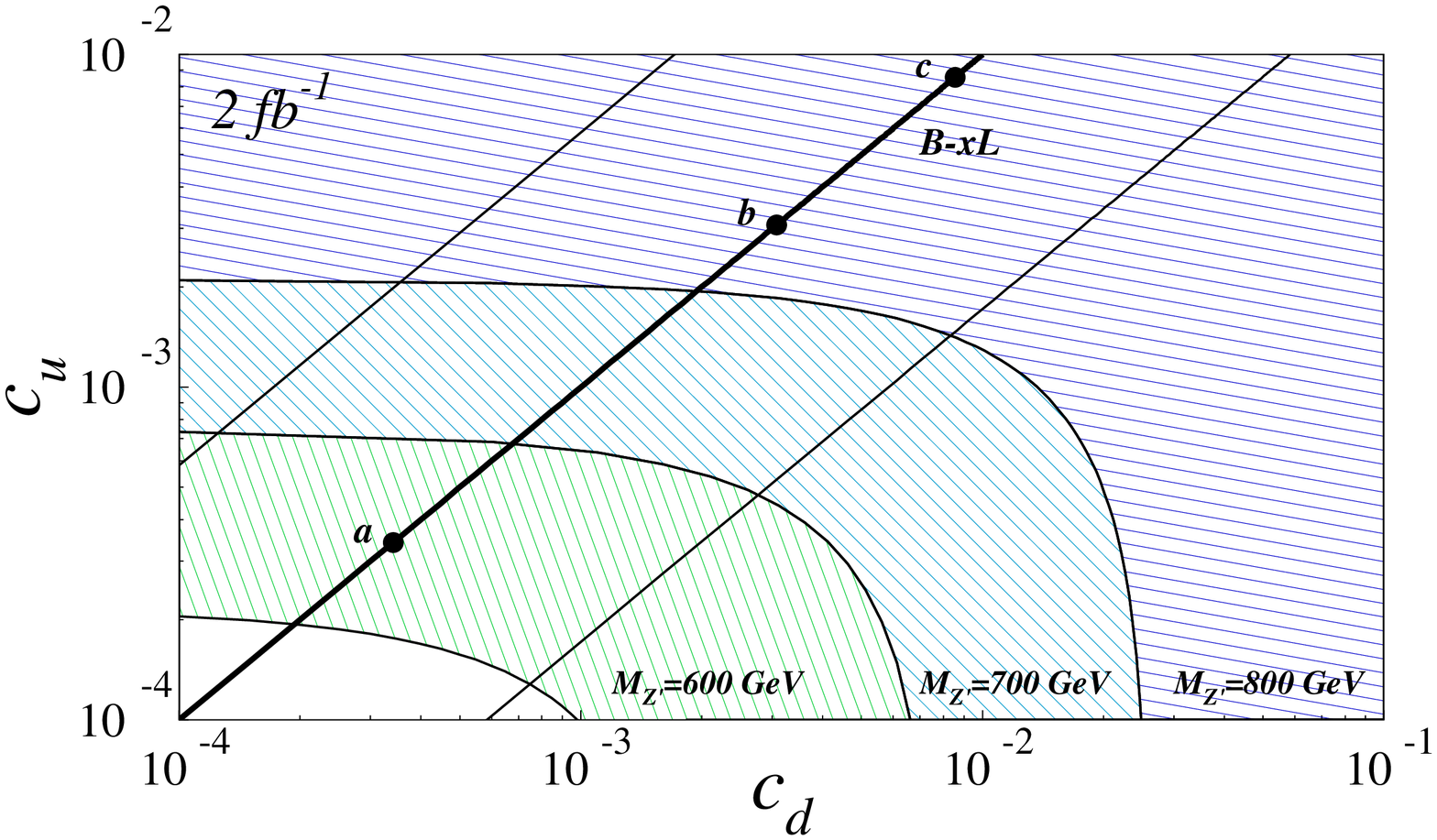}
\end{center}
\caption{Projected bounds for the $B-xL$ (left upper panel) and 
$10+x\bar{5}$ (right upper panel) models and projected excluded
regions in the $c_d-c_u$ plane (lower panel). In the two upper
panels, the vertical marks show the current LEP bounds 
for the $Z^\prime$ mass obtained as described 
in Section 2.3. In the $B-xL$ case, for $x=3$ this last bound is
$M_{Z^\prime}\ge 1800\,\mbox{GeV}$; for the 
$10+x\bar{5}$, the bound for $x=0$ is $M_{Z^\prime}\ge 119\,\mbox{GeV}$,
beyond the scope of the figure. 
The projected bounds at 
luminosities ${\cal L}=2\,\mbox{fb}^{-1}\,,\,\,10\,\mbox{fb}^{-1}$ are 
obtained from Fig.~2
by scaling with a factor $1/\sqrt{\cal L}$. The lower panel also shows
the regions in the $c_d-c_u$ plane corresponding to the $B-xL$ and $q+x u$
models, as in Figure~\protect{\ref{fig:exclcucd}}, showing the projected
increase in reach with 2 ${\rm fb}^{-1}$. 
The dots labeled $a$, $b$ and $c$ correspond to the $B-L$ model with
 $g_z=0.1$, $g_z=0.3$ and $g_z=0.5$, respectively.
}\label{fig:future}
\end{figure}

\subsection{Probes of $Z^\prime$ models at the Tevatron}

With the forthcoming data, the Tevatron experiments will be able to 
explore a new region in masses and couplings for many models containing 
extra neutral bosons, with the chance of making a $Z^\prime$ discovery.

In case that no signal excess is observed, the extracted limits to the 
$Z^\prime$ cross section will set new bounds on masses and couplings, or
alternatively on the $c_u$ and $c_d$ parameters discussed in previous 
sections. As an illustration, in Figure \ref{fig:future}, we show estimates
for the experimental limits, for different values of the integrated 
luminosity. For this estimates, we considered the current CDF bound  
\cite{CFD:CL} and assumed that the limit scales as the inverse of the
square root of luminosity, as will be the case provided the limit is
dominated by statistics and not systematics.
In the two upper panels, we plot the bounds on cross section
times branching ratio, as a function of $M_{Z^\prime}$, together with 
the predictions for different values of $x$ in 
the $B-xL$ (left) and in the $10+x\bar{5}$ (right) models. We also show 
the mass bounds, for the different cases, set by the LEP contact 
interaction constraints discussed before. The lower panel shows the 
excluded regions in the $c_d-c_u$ plane for different values of the $Z^\prime$
mass, assuming an integrated luminosity of $2\,\mbox{fb}^{-1}$. 

The results in Figure 
\ref{fig:future} show that, both in the cases of the $B-xL$ and
$10+x\bar{5}$ models, there is a sizable unexplored region in parameter 
space that the Tevatron will certainly be able to probe. For the $B-xL$ models,
LEP bounds are stronger for larger values of $|x|$, while the Tevatron can do much 
better for $|x| \lae 1$. For the $10+x\bar{5}$ models, the LEP bounds
are slightly weaker than in the previous case and the unbounded region in $x$ 
space is larger.

If a signal excess is seen in the invariant mass distribution, 
it would also be possible to shed some light on the nature of the
couplings of the new boson to the fermions by studying the angular
distribution of the final state leptons. As shown in the previous section, 
there are substantial differences, between models, in the predicted angular 
distribution.  
\begin{figure}[ht]
\begin{center}
\includegraphics[width=12cm]{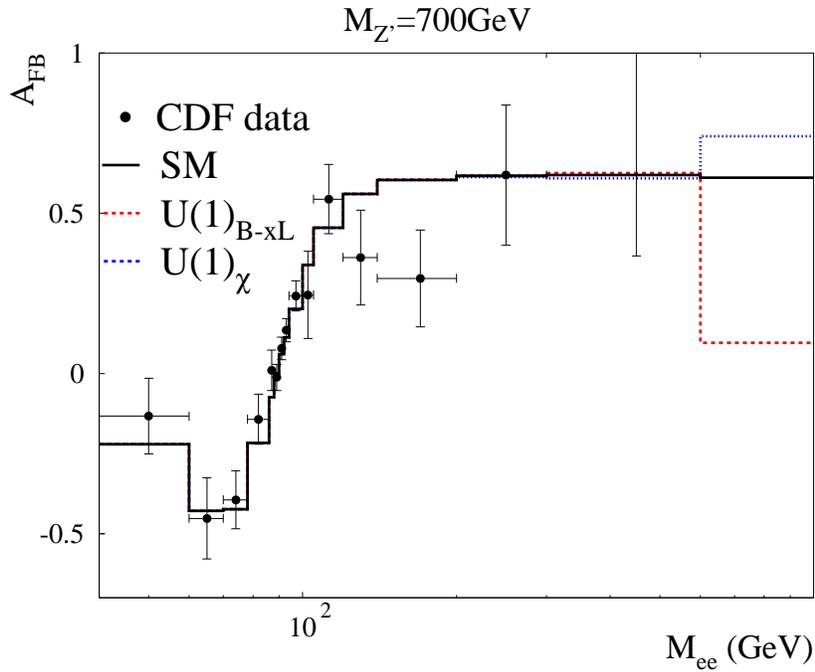}
\end{center}
\caption{The forward-backward asymmetry at LO for the $B-L$ and $U(1)_{d-xu}$
models with $M_{Z^\prime}=700\mbox{ GeV}$. 
In the case of $B-L$ we chose $g_{z}=0.1$. For the 
$U(1)_{d-xu}$ models, we fixed $x=1$ and $g_{z}=0.5$
We also show the SM prediction and 
recent CDF data for this observable \cite{CFD:FB}. 
}\label{fig:FB-TEV}
\end{figure}

In Figure \ref{fig:FB-TEV} we show the forward backward asymmetry at the 
Tevatron, predicted in the $B-xL$ and $d-xu$ models for the case of a $Z^\prime$
boson with $M_{Z^\prime}=700\,\mbox{GeV}$. The shift from the SM prediction
of the forward backward asymmetry in these two models has opposite sign, 
allowing, in principle, to distinguish between them, provided enough 
statistics are collected in the high mass region. 
Observables at the Tevatron 
other than the forward-backward asymmetry are discussed in 
Ref.~\cite{Ambrosanio:1998zf}, while capabilities of the LHC and a 
high energy $e^+ e^-$ linear collider are addressed in
Ref~\cite{Freitas:2004hq}.

\section{Conclusions}
\label{sec:conclusions}

At the Tevatron, the hypothetical $Z^\prime$ bosons may be produced via
their couplings to light quarks, and are more likely to be detected if they
decay into charged leptons. The typical signature of a $Z^\prime$
boson would be a bump in the total cross section for dilepton production
as a function of the dilepton invariant mass.
The observability of a dilepton signal in the inclusive process
$p\bar{p} \rightarrow l^+l^- X$, where $l$ stands for $e$, $\mu$ or $\tau$,
is controlled primarily by two quantities:  the $Z^\prime$ mass,
and the $Z^\prime$ decay branching fraction into $l^+l^-$ times
the hadronic structure function $W_{Z^\prime}$ defined in 
Eq.~(\ref{eq:hsf}).
The exclusion limits presented by the D0 \cite{Abachi:1996ud}-\cite{Abazov:2001qd} 
and CDF Collaborations \cite{Abe:1991vn}-\cite{Abe:1997fd} 
are curves in the plane spanned by these two parameters.
Such an exclusion plot is very useful, allowing one to derive
the range of $Z^\prime$ parameters consistent with the experiment.
However, the hadronic structure function entangles
the model dependence contained in the quark-$Z^\prime$ couplings
with the information about the proton and antiproton structures
contained in the PDFs.
In order to simplify the derivation of the exclusion limits in the
large $Z^\prime$ parameter space, we are advocating the presentation
of the exclusion curve (see Figure 3) in the $c_u-c_d$ plane,
where $c_u$ and $c_d$ are
the decay branching fraction into leptons times the
average square coupling to up and down quarks, respectively.

Assuming that the couplings of $Z^\prime$ to quarks and leptons are 
independent
of the fermion generation, the $Z^\prime$ properties are described 
primarily
by seven parameters: mass ($M_{Z^\prime}$), total width 
($\Gamma_{Z^\prime}$), and
five fermion couplings ($z_e, z_l, z_q, z_u, z_d$)$\times g_z$.
The exclusion curve in the $c_u-c_d$
plane sets a bound on a single combination of these seven parameters.
Nevertheless, in any specific model defined by certain
fermion charges it is straightforward to compute
$c_u$ and $c_d$, and to derive what is the limit on the gauge coupling $g_z$
as a function of $M_{Z^\prime}$.

For example, if the quark and lepton masses are generated by Yukawa 
couplings
to one or two Higgs doublets, as in the SM or its supersymmetric versions,
the only gauge groups that may provide a $Z^\prime$ gauge boson 
accessible at the Tevatron
are of the type $U(1)_{B-xL}$. This means that all  fermion charges are 
determined by a
single parameter, $x$. Within this family of gauge groups, $c_u$ and 
$c_d$ have a simple
dependence on $x$ and $g_z$; for a given $x$ and $M_{Z^\prime}$, the 
limit on
$g_z$ can be immediately derived.

If the quark and lepton masses are generated by a more general mechanism,
$Z^\prime$ gauge bosons associated with gauge groups other than 
$U(1)_{B-xL}$ may
be accessible at the Tevatron. We have presented three other examples of 
one-parameter
families of $U(1)$ gauge groups, chosen to include (for particular 
values of the parameter $x$)
many of the $Z^\prime$ models discussed in the literature. 
For these families of models, the Tevatron reach goes significantly
beyond the LEP II bounds
for large regions of the
three dimensional parameter space spanned by $M_{Z^\prime}$, $g_z$ and $x$.

Relaxing the assumption that the couplings of $Z^\prime$ to leptons are
generation independent, for each of the $e^+e^-$,
$\mu^+ \mu^-$ and $\tau^+ \tau^-$ final states there is a different 
$c_u$ and $c_d$.
Interestingly, $U(1)$ gauge groups that lead to a $Z^\prime$ of this type
exist even when the anomalies cancel without need for new fermions 
charged under the SM group,
and the quark and lepton masses are generated by Yukawa couplings to a 
single Higgs doublet.
Such $Z^\prime$ bosons may have very small couplings to electrons, 
evading altogether the LEP
bounds, and could be discovered in the $\mu^+ \mu^-$ or $\tau^+ \tau^-$ 
channels at the Tevatron.

Although generation-independent $Z^\prime$ couplings to quarks are 
tightly constrained by
measurements of various flavor-changing neutral currents, a $Z^\prime$ 
with different
couplings to the $d$ and $s$ quarks (or to the  $u$ and $c$ quarks)
in the mass range accessible at the Tevatron cannot be completely ruled 
out. In that case,
the $c_u$ and $c_d$ parameterization would have to be supplemented by $c_s$
(or $c_c$, $c_b$) quantities.  Even in this case, the fact that current- and
next-generation hadron colliders collide nucleons (and anti-nucleons) implies
that $c_u$ and $c_d$ typically remain the most important, because of their
large valence distribution functions (particularly at large parton $x$)
for nucleons.

Observables other than the total cross section for dilepton production
can also be measured at the Tevatron. We have discussed the additional
information provided by the forward-backward asymmetry. In most cases 
however,
a $Z^\prime$ discovery is more likely to occur first as a bump in the 
dilepton
total cross section. If that happens, $M_{Z^\prime}$ can be determined 
by the invariant mass
of the lepton pair, and a curve (actually a band of experimental error bars)
in the $c_u-c_d$ plane can be derived.
For each $Z^\prime$ model (fixed $x$ within the one-parameter families), 
the curve
would determine the gauge coupling.
However, pinning down the model would be difficult, requiring additional 
observables
at the Tevatron and future colliders.

\bigskip\bigskip

{\bf Acknowledgements}: 

The authors have benefitted from discussions with Ayres Freitas and
Beate Heinemann. A.D. thanks the Theory Department at Fermilab for their
warm hospitality and financial support, and CONICET, Argentina, for financial 
support.
Fermilab is operated by Universities Research Association Inc.  under  
contract no. DE-AC02-76CH02000 with the DOE.


\section*{Appendix: Interference terms}
\setcounter{equation}{0}

The $Z^\prime$ interference with the $Z$ and the photon is taken into account 
by the second term in Eq.~(\ref{eq:cs1}). This term can be 
factorized similarly to the contribution due solely to the
$Z^\prime$ [first term in Eq.~(\ref{eq:cs1}], by replacing 
$\sigma(Z^\prime\rightarrow l^+l^-)$ 
with
\begin{equation}\label{eq:slepinterf}
\sigma(Z^\prime,X)
=\frac{g_z g_X}{2\pi}\,
\left( \frac{ z_{l_j}\,z^{X}_{l} + z_{e_j}\,z^{X}_{e} }{288}\right)
\,\frac{(Q^2-M_{Z^\prime}^2)\,(Q^2-M_{X}^2)+M_{Z^\prime}\,M_X\,
\Gamma_{Z^\prime}\,\Gamma_{X}}
{
\left[ \left( Q^2-M_{Z^{\prime}}^2 \right)^2
+ M_{Z^{\prime}}^2\,\Gamma_{Z^{\prime}}^2\right]\,
\left[ \left( Q^2-M_{X}^2 \right)^2 + M_{X}\,\Gamma_{X}^2 \right]
}\,, \nonumber
\end{equation}
where $X=\gamma, Z^{0}$ and $g_X$, $(z^{X}_{l_j},z^{X}_{e_j})$, 
$M_X$ and $\Gamma_X$ are the corresponding coupling, lepton charges and
mass and width of the boson. The quark charges in 
$W_{Z^{\prime}}$ must also be changed accordingly in Eq.~(\ref{eq:cs1}).

\begin{figure}[t]
\begin{center}
\begin{minipage}{7.9cm}
\begin{center}
\includegraphics[width=8.2cm]{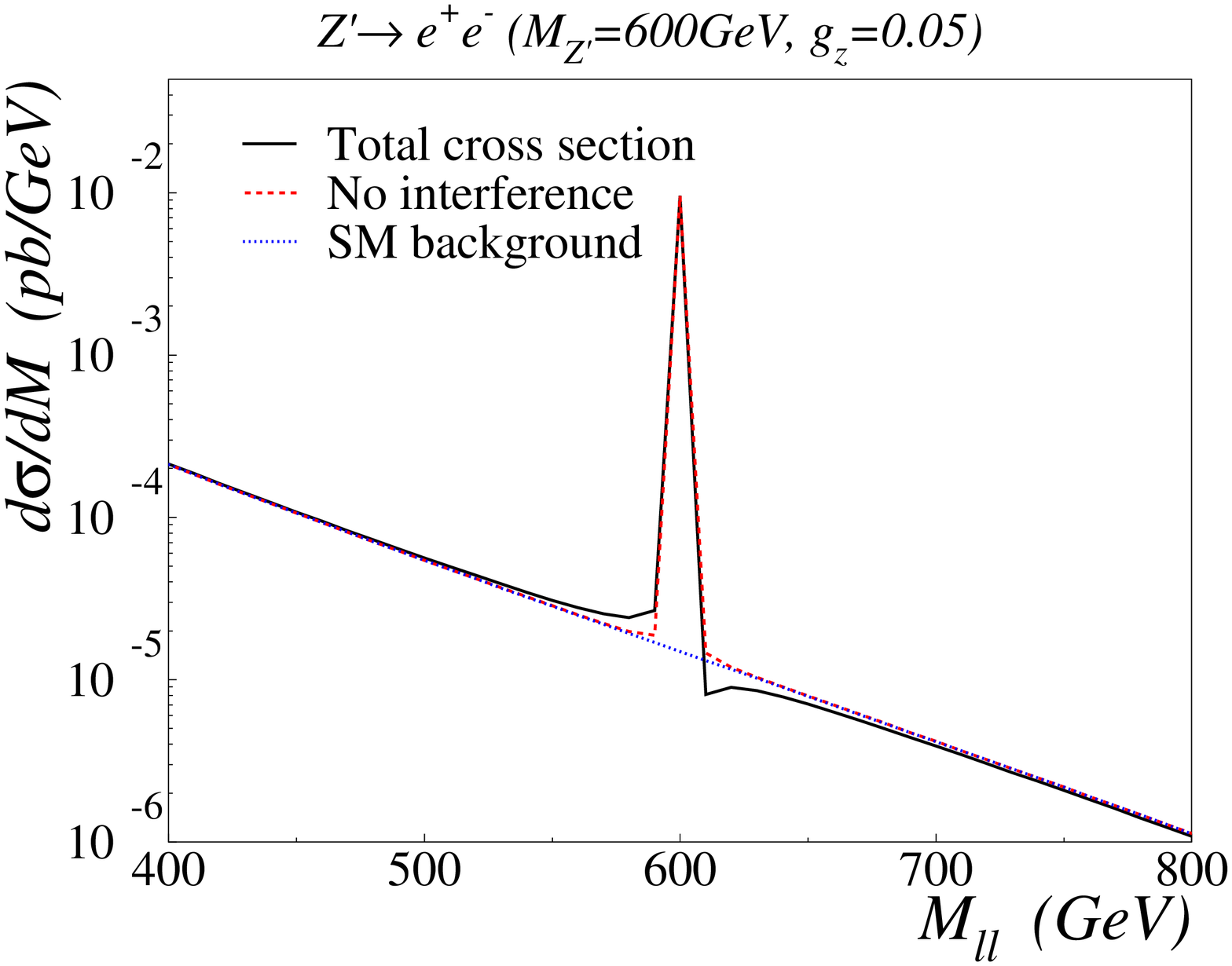}
{\bf (a)}
\end{center}
\end{minipage}
\begin{minipage}{7.9cm}
\begin{center}
\includegraphics[width=8.2cm]{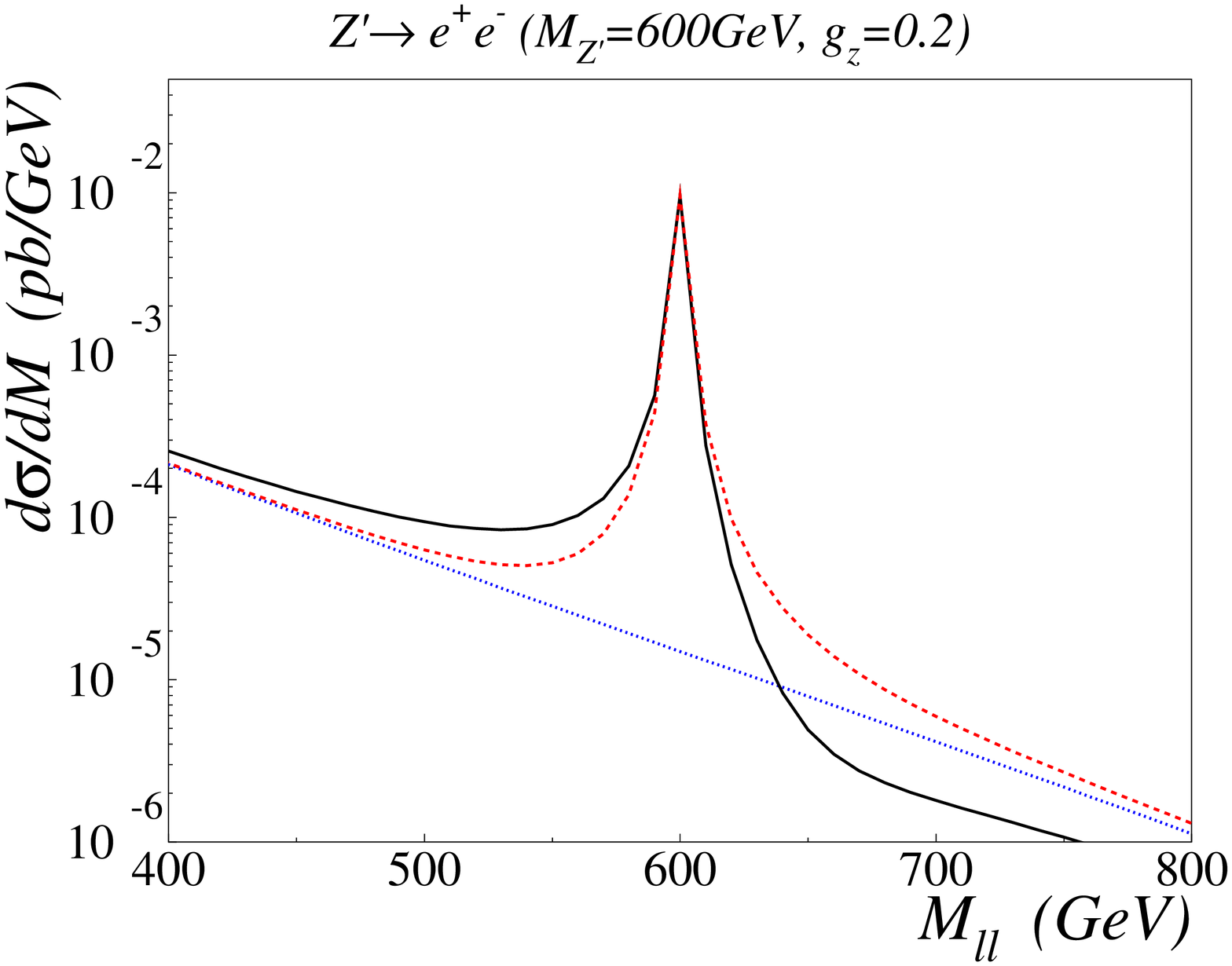}
{\bf (b)}
\end{center}
\end{minipage}
\end{center}
\caption{NLO differential cross sections for the production of 
electron-positron pairs as a function of the invariant mass of the pair, 
in the $U(1)_{B+L}$ model. The solid curves correspond to the total cross
sections for the signal, including interference terms, for the gauge coupling
fixed to $g_z=0.05$ and $g_z=0.2$ respectively. 
Dashed lines are the same cross 
sections neglecting interference terms and the dotted lines correspond to the
SM background.}
\label{fig:interf}
\end{figure}

For a  narrow $Z^\prime$ resonance,
the interference of the $Z^\prime$ with
the $Z$ and photons can be neglected. As an illustration, in Figure 
\ref{fig:interf}, we plot the NLO cross section for the production of an 
electron-positron pair as a function of the lepton system invariant mass.
The curves shown correspond to the SM background for the process and to
the $Z^\prime$ mediated ones, with and without the interference terms with 
the $Z^{0}$ and the photon. For the plot we chose the $B-xL$ model, discussed
in the previous section, and fixed $M_{Z^{\prime}}=600\mbox{ GeV}$, 
and $x=-1$. Note that for $x=+1$ the only difference is that the
interference terms change sign. The two sets 
of curves correspond to different choices for the gauge coupling, namely
$g_{z}=0.05$ and $g_{z}=0.2$.
These values of the coupling correspond to $\Gamma_{Z^\prime} = 0.26\mbox{ GeV}$ and
$\Gamma_{Z^\prime} = 4.15\mbox{ GeV}$, respectively, 
assuming that only decays to SM particles are allowed and neglecting all QCD and
electroweak corrections. The bounds set by LEP for these two cases are
$M_{Z^\prime}\ge 300,\, 1200 \mbox{ GeV}$ respectively. 
In the case of $g_{z}=0.05$, 
the signal cross section outside the $Z^\prime$ peak is completely 
negligible compared to the SM background, and, at the peak, the interference 
terms can be neglected. For $g_{z}=0.2$, the interference terms 
are more important and contribute to the tails at low and high mass. 
However, the experimental errors would not allow one to disentangle the 
signal from the background outside the peak, where again, the
signal cross section is dominated by terms containing only $Z^\prime$ 
propagators.


\vfil

\begin{thebibliography}{99} \frenchspacing

\bibitem{Eichten:1984eu}
E.~Eichten, I.~Hinchliffe, K.~D.~Lane and C.~Quigg,
``Super Collider Physics,''
Rev.\ Mod.\ Phys.\  {\bf 56}, 579 (1984)
[Addendum-ibid.\  {\bf 58}, 1065 (1986)].
 

\bibitem{Hewett:1988xc} 
For a review, see
J.~L.~Hewett and T.~G.~Rizzo, ``Low-Energy Phenomenology Of Superstring
Inspired E(6) Models,''
Phys.\ Rept.\  {\bf 183}, 193 (1989); 

\bibitem{Cvetic:1995rj} 
See, {\it e.g.},
M.~Cvetic and P.~Langacker, ``Implications of Abelian Extended Gauge
Structures From String Models,'' Phys.\ Rev.\ D {\bf 54}, 3570 (1996)
[arXiv:hep-ph/9511378].

\bibitem{Leike:1998wr}
For a review, see A.~Leike,
``The phenomenology of extra neutral gauge bosons,''
Phys.\ Rept.\  {\bf 317}, 143 (1999)
[arXiv:hep-ph/9805494].

\bibitem{Appelquist:2002mw}
T.~Appelquist, B.~A.~Dobrescu and A.~R.~Hopper,
``Nonexotic neutral gauge bosons,''
Phys.\ Rev.\ D {\bf 68}, 035012 (2003)
[arXiv:hep-ph/0212073].

\bibitem{Freitas:2004hq}
A.~Freitas,
``Weakly coupled neutral gauge bosons at future linear colliders,''
arXiv:hep-ph/0403288.

\bibitem{Altarelli:2004fq}
For a recent review, see,
G.~Altarelli and M.~W.~Grunewald,
``Precision electroweak tests of the standard model,''
arXiv:hep-ph/0404165.

\bibitem{Abreu:1994ri}
P.~Abreu {\it et al.}  [DELPHI Collaboration],
Z.\ Phys.\ C {\bf 65}, 603 (1995).

\bibitem{Chanowitz:2001bv}
M.~S.~Chanowitz,
``The Z $\to$ anti-b b decay asymmetry: Lose-lose for the standard model,''
Phys.\ Rev.\ Lett.\  {\bf 87}, 231802 (2001)
[arXiv:hep-ph/0104024];
M.~S.~Chanowitz,
``Electroweak data and the Higgs boson mass: A case for new physics,''
Phys.\ Rev.\ D {\bf 66}, 073002 (2002)
[arXiv:hep-ph/0207123];
D.~Choudhury, T.~M.~P.~Tait and C.~E.~M.~Wagner,
``Beautiful mirrors and precision electroweak data,''
Phys.\ Rev.\ D {\bf 65}, 053002 (2002)
[arXiv:hep-ph/0109097].

\bibitem{Erler:1999nx}
J.~Erler and P.~Langacker,
``Indications for an extra neutral gauge boson in electroweak precision
data,''
Phys.\ Rev.\ Lett.\  {\bf 84}, 212 (2000)
[arXiv:hep-ph/9910315];
P.~Langacker and M.~Plumacher,
``Flavor changing effects in theories with a heavy Z' boson with family
non-universal couplings,''
Phys.\ Rev.\ D {\bf 62}, 013006 (2000)
[arXiv:hep-ph/0001204].

\bibitem{Carena:2002me}
M.~Carena, T.~M.~P.~Tait and C.~E.~M.~Wagner,
``Branes and orbifolds are opaque,''
Acta Phys.\ Polon.\ B {\bf 33}, 2355 (2002)
[arXiv:hep-ph/0207056].

\bibitem{Arkani-Hamed:2002qy}
For example,
N.~Arkani-Hamed, A.~G.~Cohen, E.~Katz and A.~E.~Nelson,
``The littlest Higgs,''
JHEP {\bf 0207}, 034 (2002)
[arXiv:hep-ph/0206021].

\bibitem{Csaki:2002qg}
C.~Csaki, J.~Hubisz, G.~D.~Kribs, P.~Meade and J.~Terning,
``Big corrections from a little Higgs,''
Phys.\ Rev.\ D {\bf 67}, 115002 (2003)
[arXiv:hep-ph/0211124];
J.~L.~Hewett, F.~J.~Petriello and T.~G.~Rizzo,
``Constraining the littlest Higgs. ((U)),''
JHEP {\bf 0310}, 062 (2003)
[arXiv:hep-ph/0211218].

\bibitem{lep:2003ih}
  [LEP Collaboration],
``A combination of preliminary electroweak measurements and constraints on the
standard model,''
arXiv:hep-ex/0312023.

\bibitem{Hill:1994hp}
C.~T.~Hill,
``Topcolor assisted technicolor,''
Phys.\ Lett.\ B {\bf 345}, 483 (1995)
[arXiv:hep-ph/9411426];
R.~S.~Chivukula and E.~H.~Simmons,
``Electroweak limits on non-universal Z' bosons,''
Phys.\ Rev.\ D {\bf 66}, 015006 (2002)
[arXiv:hep-ph/0205064].

\bibitem{Chivukula:1995gu}  
R.~S.~Chivukula, E.~H.~Simmons and J.~Terning,  
``Limits on noncommuting extended technicolor,''  
Phys.\ Rev.\ D {\bf 53}, 5258 (1996)  
[arXiv:hep-ph/9506427]; 
D.~J.~Muller and S.~Nandi,  
``Topflavor: A Separate SU(2) for the Third Family,''  
Phys.\ Lett.\ B {\bf 383}, 345 (1996)  
[arXiv:hep-ph/9602390];  
E.~Malkawi, T.~Tait and C.~P.~Yuan,  
``A Model of Strong Flavor Dynamics for the Top Quark,''  
Phys.\ Lett.\ B {\bf 385}, 304 (1996)  
[arXiv:hep-ph/9603349]; 
H.~J.~He, T.~Tait and C.~P.~Yuan, 
``New topflavor models with seesaw mechanism,'' 
Phys.\ Rev.\ D {\bf 62}, 011702 (2000) 
[arXiv:hep-ph/9911266]. 

\bibitem{Batra:2003nj}
P.~Batra, A.~Delgado, D.~E.~Kaplan and T.~M.~P.~Tait,
``The Higgs mass bound in gauge extensions of the minimal supersymmetric
standard model,''
JHEP {\bf 0402}, 043 (2004)
[arXiv:hep-ph/0309149];
P.~Batra, A.~Delgado, D.~E.~Kaplan and T.~M.~P.~Tait,
``Running into new territory in SUSY parameter space,''
arXiv:hep-ph/0404251.

\bibitem{Abe:1991vn}
F.~Abe {\it et al.}  [CDF Collaboration],
``A Search for new gauge bosons in anti-p p collisions at S**(1/2) = 1.8-TeV,''
Phys.\ Rev.\ Lett.\  {\bf 68}, 1463 (1992).

\bibitem{Abe:1994ns}
F.~Abe {\it et al.}  [CDF Collaboration],
``Search for new gauge bosons decaying into dielectrons in anti-p p collisions at s**(1/2) 1.8-TeV,''
Phys.\ Rev.\ D {\bf 51}, 949 (1995).

\bibitem{Abe:1997fd}
F.~Abe {\it et al.}  [CDF Collaboration],
``Search for new gauge bosons decaying into dileptons in anti-p p  collisions at s**(1/2) = 1.8-TeV,''
Phys.\ Rev.\ Lett.\  {\bf 79}, 2192 (1997).

\bibitem{Abachi:1996ud}
S.~Abachi {\it et al.}  [D0 Collaboration],
``Search for additional neutral gauge bosons,''
Phys.\ Lett.\ B {\bf 385}, 471 (1996).

\bibitem{Abbott:1998rr}
B.~Abbott {\it et al.}  [D0 Collaboration],
``Measurement of the high-mass Drell-Yan cross section and limits on
quark-electron compositeness scales,''
Phys.\ Rev.\ Lett.\  {\bf 82}, 4769 (1999)
[arXiv:hep-ex/9812010].

\bibitem{Abazov:2001qd}
V.~M.~Abazov {\it et al.}  [D0 Collaboration],
``Search for heavy particles decaying into electron positron pairs in p  anti-p
collisions,''
Phys.\ Rev.\ Lett.\  {\bf 87}, 061802 (2001)
[arXiv:hep-ex/0102048].

\bibitem{FNALJET}
Talk at Fermilab by Anton Anastassov, July 23, 2004, 
http://theory.fnal.gov/jetp/previous.html.

\bibitem{Hamberg:1990np}
R.~Hamberg, W.~L.~van Neerven and T.~Matsuura,
``A Complete Calculation Of The Order Alpha-S**2 Correction To The Drell-Yan K
Factor,''
Nucl.\ Phys.\ B {\bf 359}, 343 (1991)
[Erratum-ibid.\ B {\bf 644}, 403 (2002)].

\bibitem{Harlander:2002wh}
R.~V.~Harlander and W.~B.~Kilgore,
``Next-to-next-to-leading order Higgs production at hadron colliders,''
Phys.\ Rev.\ Lett.\  {\bf 88}, 201801 (2002)
[arXiv:hep-ph/0201206].

\bibitem{D0:CL} D0 Collaboration, note 4375-Conf, 
``Search for heavy Z' Bosons in the Dielectron Channel with 200 pb$^{-1}$ of Data.''

\bibitem{CFD:CL}
Tracey Pratt (for the CDF Collaboration), talk at the SUSY 2004
Conference, June 2004. 

\bibitem{Martin:2002dr}
A.~D.~Martin, R.~G.~Roberts, W.~J.~Stirling and R.~S.~Thorne,
``NNLO global parton analysis,''
Phys.\ Lett.\ B {\bf 531}, 216 (2002)
[arXiv:hep-ph/0201127].

\bibitem{Pumplin:2002vw}
J.~Pumplin, D.~R.~Stump, J.~Huston, H.~L.~Lai, P.~Nadolsky and W.~K.~Tung,
JHEP {\bf 0207}, 012 (2002)
[arXiv:hep-ph/0201195].

\bibitem{vNeerven:zwprod}
http://www.lorentz.leidenuniv.nl/~neerven/

\bibitem{Baur:2001ze}
U.~Baur, O.~Brein, W.~Hollik, C.~Schappacher and D.~Wackeroth,
``Electroweak radiative corrections to neutral-current Drell-Yan  processes at
hadron colliders,''
Phys.\ Rev.\ D {\bf 65}, 033007 (2002)
[arXiv:hep-ph/0108274].

\bibitem{Ciafaloni:1998xg}
P.~Ciafaloni and D.~Comelli,
``Sudakov effects in electroweak corrections,''
Phys.\ Lett.\ B {\bf 446}, 278 (1999)
[arXiv:hep-ph/9809321].

\bibitem{Baur:1997wa}
U.~Baur, S.~Keller and W.~K.~Sakumoto,
``QED radiative corrections to Z boson production and the forward  backward
Phys.\ Rev.\ D {\bf 57}, 199 (1998)
[arXiv:hep-ph/9707301].

\bibitem{CFD:FB}
Data extracted from the plots in
http://www-cdf.fnal.gov/physics/ewk/2004/afb/

\bibitem{Ambrosanio:1998zf}
S.~Ambrosanio {\it et al.}  [MSSM Working Group Collaboration],
``Report of the Beyond the MSSM subgroup for the Tevatron Run II SUSY /  Higgs
workshop,'' Sec. XV
arXiv:hep-ph/0006162.




\end{thebibliography}
\end{document}